\ifdefined\pdfoutput\pdfoutput=1\fi
\documentclass[journal]{IEEEtran}
\usepackage{amsmath,amsfonts}
\usepackage[style=ieee, backend=biber,giveninits=true, style=numeric, maxnames=99, sorting=none, isbn=false]{biblatex}
\AtEveryBibitem{\clearfield{month}}
\usepackage{booktabs}
\usepackage{colortbl} 
\usepackage{multirow}
\usepackage[most]{tcolorbox}
\usepackage{dsfont}
\usepackage{command_MU}
\usepackage{amsmath}
\usepackage{scalerel}
\usepackage{algorithm}
\usepackage{algpseudocode}
\usepackage{graphicx}
\usepackage{tikz}
\usetikzlibrary{positioning}
\usetikzlibrary{calc}
\usetikzlibrary{shapes,arrows}
\usepackage{algpseudocode}
\usepackage{algorithm}
\usepackage{stmaryrd}
\usepackage{tikz}
\usetikzlibrary{calc, arrows.meta}
\usepackage[outline]{contour}
\usepackage{dsfont}

\usepackage{hyperref}
\usepackage{placeins}
\usepackage{pgfplots}
\pgfplotsset{compat=newest}
\usetikzlibrary{fit}

\newcommand{\vo}{{\mathbf{s}}}
\newcommand{\vd}{{\boldsymbol{\theta}}}
\newcommand{\vk}{\mathbf{k}}
\newcommand{\vx}{\mathbf{x}}
\newcommand{\vy}{\mathbf{y}}
\newcommand{\vn}{\mathbf{n}}
\newcommand{\vA}{\mathbf{A}}
\newcommand{\Proj}[1]{\mathcal{P}#1}
\usepackage{bm}

\usepackage[most]{tcolorbox}
\definecolor{softblue}{rgb}{0.0627, 0.3333, 0.6039} 
\hypersetup{
    colorlinks=true,        
    linkcolor=softblue,        
    citecolor=softblue,        
    urlcolor=softblue,         
    pdfborder={0 0 0}      
}

\usepackage{enumitem}

\DeclareMathOperator*{\argmin}{arg\,min}

\pgfplotsset{compat=1.18}

\definecolor{tabblue}{rgb}{0.1215, 0.4666, 0.7058}
\definecolor{taborange}{rgb}{1.0, 0.4980, 0.0549}

\title{Geometry Calibration in Tomography \\ with a Differentiable Ray-Based Model}

\author{
Youssef Haouchat, Aleix Boquet-Pujadas, Sepand Kashani, Philippe Thévenaz,~\IEEEmembership{Fellow, IEEE},\\ and Michael Unser,~\IEEEmembership{Life Fellow, IEEE}
\vspace{-24.7pt}
\thanks{Y. Haouchat, P. Thévenaz, and M. Unser are with the Biomedical Imaging Group at EPFL, Switzerland. 
(email: youssef.haouchat@epfl.ch). 
A. Boquet-Pujadas is with the SDSC hub at PSI, Switzerland. 
(email: aleix.boquet-i-pujadas@psi.ch). 
S. Kashani is with the Center for Imaging at EPFL, Switzerland. Corresponding author: A. Boquet-Pujadas. 
This work was supported in part by the European Research Council (ERC-2020-AdG FunLearn-101020573) and by the Swiss National Science Foundation (Sinergia Grant CRSII5 198569).
}
}

\AtEveryBibitem{\clearfield{doi} \clearfield{url} \clearfield{eprint}}

\begin{document}

\newcommand\submittedtext{%
  \footnotesize This work has been submitted to the IEEE for possible publication. Copyright may be transferred without notice, after which this version may no longer be accessible.}

\newcommand\submittednotice{%
  \begin{tikzpicture}[remember picture,overlay]
    \node[anchor=south,yshift=10pt] at (current page.south) {\fbox{\parbox{\dimexpr0.65\textwidth-\fboxsep-\fboxrule\relax}{\submittedtext}}};
  \end{tikzpicture}%
}
\let\nablaorig\nabla
\renewcommand{\nabla}{\bm{\nablaorig}}

 \maketitle
\submittednotice

\begin{abstract}
Geometric misalignments between the nominal and true acquisition parameters in tomography degrade reconstructions. 
We propose a framework that jointly reconstructs the volume and calibrates the acquisition geometry for arbitrary source--detector configurations.
The core of our framework is an x-ray transform operator whose gradients with respect to the acquisition geometry can be efficiently computed with a ray-tracing method of structure and computational complexity similar to those of the forward operator. We represent the volume in a B-spline basis to provide a continuously differentiable model. This results in a better-behaved optimization landscape compared to voxel-based representations. 
We validate our framework with CT, micro-CT, nano-CT, and positron emission tomography data under a variety of geometric misalignments. 
\end{abstract}


\input{fig_illustration}

\section{Introduction}
\subsection{Motivation and Related Work}
Tomography is an imaging modality where the goal is to reconstruct a volume from a set of its line integrals. The intercepts and directions of these lines constitute the \textit{acquisition parameters}.
While standard reconstruction methods typically assume their perfect calibration, the acquisition parameters of real-world imaging systems such as computed tomography (CT) scanners often deviate from their nominal values. This issue has become more significant with the advent of microscopic and nanoscopic tomography~\cite{nanoscale}, where mechanical instabilities are large relative to the spatial resolution of the system. Common sources of geometric mismatches in standard tomographic setups include motor imprecision, mechanical vibrations, axis misalignment, and focal-spot instability. They often manifest themselves as angular errors, detector displacements, geometric precession, and source-position shifts (Figure~\ref{fig:calibration_scenarios}). It is important to account for these mismatches because they introduce artifacts that can significantly degrade the quality of reconstructed volumes~\cite{cone_beam_misalignment}. 
The most widely-adopted methods 
to mitigate these  
errors still rely on physical calibration objects~\cite{calibration_objects} or sinogram-based corrections, common in CT or electron microscopy. They include: (i) fiducial-aided methods that track high-contrast markers~\cite{fiducials}, (ii) feature-matching techniques such as center-of-mass tracking~\cite{COM} or SIFT~\cite{sift}, and (iii) iterative methods such as outer-contour-based misalignment correction~\cite{ocmc}. Among approaches that employ fiducial markers, positron emission tomography (PET) scanners, for example, can be calibrated by triangulating a rotating point source~\cite{pierce2012detector}. 
However, all these methods are tailored to particular calibration tasks and cannot be readily adapted to arbitrary acquisition geometries.

    The gradient-based optimization of the acquisition parameters would enable the calibration of arbitrary geometric configurations, but this is conditioned on the availability of a forward model that is efficiently differentiable with respect to those parameters. Existing 
    open-source libraries \cite{vanAarle2015Astra, biguri2016tigre, gursoy2014tomopy} usually lack support for gradients with respect to acquisition geometry. The implementation of these gradients is challenging. While automatic-differentiation-based projectors support geometric gradients, they often incur a high memory overhead and a reduced computational efficiency compared to dedicated ray-tracing kernels, as explained in Section \ref{sec:implementation}.
    Likewise, alternative methods relying on finite-difference approximations \cite{finite-diff} are computationally expensive and do not scale well to high-dimensional parameter spaces. 
    
    Recent work has achieved calibration with implicit neural representation \cite{dokmanic} where a neural network models the x-ray measurement operator, thus naturally supports gradient computation with respect to its input acquisition parameters. Others achieved explicit gradient computation~\cite{Liu_optical_tomography, Zehni_gradient_cryo}, but these approaches perform exhaustive voxel-by-voxel calculations, which is resource-intensive.

    We propose an augmented variant of ray-tracing that computes explicitly the exact gradient of the x-ray transform with respect to arbitrary acquisition parameters.
    By representing the reconstructed volume with differentiable basis functions, we ensure that the forward model is continuously differentiable with respect to the geometry, which improves convergence during optimization. Our contributions enable a wide range of calibration scenarios to be handled effectively within a unified framework. We validate this framework using both real and simulated data from CT, micro-CT, nano-CT, and PET.

    \subsection{X-Ray Model and Ray Tracing}
In tomography, measurements are assumed to be line integrals of the function that describes the volume of interest.
Let a line in $\R^3$ be parameterized in terms of $\vo$ and $\vd$ as
\begin{equation}
\label{rayparam}
    \{\vo + t\vd \in \mathbb{R}^3 \mid t \in \mathbb{R}\},
\end{equation}
where $\vd \in \mathbb{S}^2$ is the unit vector that directs the line and $\vo \in \R^3$ is its intercept.
We drop the canonical orthogonality constraint $\langle \vo, \vd\rangle=0$ to simplify our derivations and streamline the implementation.

\vspace{2pt}

The x-ray transform \cite{Kak, natterer2001mathematics} denoted by $\mathcal{P}$ of an integrable function \(\text{$f : \mathbb{R}^3 \to \mathbb{R}$}\) corresponds to the collection of all its integrals along such lines. With the current parameterization, it is expressed as \mbox{$\Proj{f}:(\R^3, \mathbb{S}^2) \rightarrow\R$}, with
\begin{equation}
\label{xrt}
    \Proj{f}(\vo, \vd) 
    = \int_{\mathbb{R}} f(\vo + t\vd) \, \mathrm{d}t.
\end{equation}
In practice, the numerical computation of \eqref{xrt} usually assumes a discretization of $f$ as a linear combination of basis functions that are translates of a generator $\varphi$ placed on a Cartesian grid. 
To simplify the presentation and without loss of generality, we assume that these functions are positioned on $\Z^3$. Accordingly, the continuous-domain image to be reconstructed is written as
\begin{equation}
\label{decomposition}
    \vx \mapsto f(\vx) = \sum_{\vk\in\Omega}c_\vk\varphi(\vx - \vk),
\end{equation}
where $\Omega = \{1, \ldots, N\}^3$ with $N^3$ the number of coefficients that describe the volume $f$, and $c_\vk$ is the coefficient associated to the basis function $\varphi(\cdot-\vk)$. The x-ray transform of $f$ is
\begin{equation}
    \label{discrete_xrt}
    \Proj{f}(\vo, \vd) = \sum_{\vk\in\Omega}c_\vk \Proj{\varphi}(\vo - \text{Proj}_{H_{\vd}}(\vk), \vd),
\end{equation}

where $\text{Proj}_{H_{\vd}}(\vk)$ is the projection of $\vk$ onto the hyperplane orthogonal to $\vd$.

For any acquisition parameters $(\vo, \vd)$, the efficient computation of 
\eqref{discrete_xrt} requires one to identify the indices $\vk \in \Omega$ 
whose associated basis function $\varphi(\cdot - \vk)$ has a nonzero integral along 
the ray, so that $\Proj{\varphi}(\vo - \text{Proj}_{H_{\vd}}(\vk), \vd) \neq 0$. 
Ray-tracing algorithms are well-suited to this purpose: they sequentially compute 
the intersection points between the ray and cells of the Cartesian grid. 
When $\varphi$ is the indicator function of the unit cube (voxel), 
$\Proj{\varphi}(\vo - \text{Proj}_{H_{\vd}}(\vk), \vd)$ equals the intersection 
length of the ray with the $\vk$th cell, so $\Proj{f}(\vo, \vd)$ reduces to a 
weighted sum of these lengths with coefficients $c_\vk$. In this case, the algorithm 
is known either as the differential digital analyser (DDA) or as Siddon's method~\cite{siddon, amanitides}.
When $\varphi$ is a higher-order spline function, \textit{generalized} DDA algorithms that can handle overlapping basis functions~\cite{haouchattci, haouchatisbi} are deployed to efficiently compute the corresponding coefficients.
However, there exists no explicit ray-tracing algorithm that provides the gradients of $\Proj{f}$ with respect 
to the acquisition parameters $(\vo, \vd)$, which are required for geometry optimization. 

\subsection{Contributions}
Our contributions are as follows.
\begin{enumerate}
    \item We formulate the gradient of $\Proj{f}$ with respect to the geometric parameters $(\vo, \vd)$ and obtain formulas that are computable with ray-tracing (Equations~\eqref{eq:grad_s} and \eqref{eq:grad_d}).
    \item We provide practical methods (Algorithms~\ref{alg1} and \ref{alg2}) to perform the task efficiently. Their basis-function-specific subroutines are obtained either in 
    closed form or via automatic differentiation of the closed-form line 
    integral alone, as we apply to higher-order splines.
    \item We use differentiable basis functions in~\eqref{decomposition} to ensure that the 
    forward model is continuously differentiable also with respect to the geometry. The structure and efficiency of the resulting algorithm follows that of ray-tracing.
    \item As a baseline, we also derive the analytical gradients for the conventional voxel model which is not differentiable in the classic sense (Algorithms~\ref{alg:vox1} and \ref{alg:vox2}).
    \item We demonstrate our method on well-known calibration scenarios and show 
    the benefit of smooth basis functions over voxels for geometry optimization.
    \item We provide an implementation that is publicly available.\footnote{\url{https://github.com/HaouchatY/differentiable_xrt}}
\end{enumerate}

\section{Method and Implementation}
We derive the formulas and algorithms that are used in our framework. By contrast with most implementations that use voxels as basis functions to describe the volume, our formulation involves differentiable basis functions. First, we introduce the general ray-tracing Algorithms~\ref{alg1} and~\ref{alg2} that provide an exact computation of the gradients. Then, we instantiate them either with differentiable basis functions (Section~\ref{sec:ad}),
or with 
explicit subroutines for voxels (Section~\ref{sec:voxels}).

    \subsection{Main Derivations}

We adopt the following notation for the gradients of the partial mappings of the x-ray transform: for $\vo\in\R^3$ and \mbox{$\vd \in \mathbb{S}^2$},
\begin{align}
    \nabla_\vo[\Proj{f}](\vo, \vd) &:= \nabla[\Proj{f}(\cdot, \vd)](\vo), \notag\\
    \nabla_\vd[\Proj{f}](\vo, \vd) &:= \nabla[\Proj{f}(\vo, \cdot)](\vd).
\end{align}
In the second mapping, $\nabla$ denotes the Riemannian gradient on the unit sphere $\mathbb{S}^2$, which corresponds to the projection of the Euclidean gradient onto the plane tangent to the sphere at $\vd$.
We assume that $f$ is compactly supported and continuously differentiable, so 
that the Leibniz rule allows differentiation under the integral sign.\footnote{When 
$\varphi$ is not classically differentiable (e.g., the voxel indicator), 
$\nabla\varphi$ is understood in the distributional sense and the interchange 
follows from the generalized Leibniz rule for distributions.} Since the integral of a vector-valued function is defined component-wise, we obtain that
\begin{align}
    \nabla_\vo[\Proj{f}](\vo, \vd) &= \int_\R 
    \nabla[\vo' \mapsto f(\vo' + t\vd)](\vo)
    \dd t \notag\\  
    &= \int_\R 
    \nabla f(\vo + t\vd)
    \dd t \notag \\ 
    &= \begin{pmatrix}
    \Proj{(\partial_1 f)}(\vo, \vd) \\
    \Proj{(\partial_2 f)}(\vo, \vd) \\
    \Proj{(\partial_3 f)}(\vo, \vd)
    \end{pmatrix}. \label{eq:vector_form} 
\end{align}

The second equality comes from the chain rule and the fact that the Jacobian of $\vo \mapsto \vo + t \vd$ is the identity of $\R^3$.
Similarly, we write that
\begin{align}
    \nabla_\vd[\Proj{f}](\vo, \vd) &= \int_\R 
    \nabla[\vd' \mapsto f(\vo + t\vd')](\vd)
    \dd t \notag\\
    &= (\mathbf{I} - \vd\vd^\top)
    \int_\R 
    t\nabla f(\vo + t\vd)
    \dd t,
\end{align}
where $t(\mathbf{I} - \vd\vd^\top)$ is the Riemannian Jacobian of $\vd \mapsto \vo + t \vd$ on $\mathbb{S}^2$. Both gradients involve $\nabla f$ under the integral or, equivalently, $\nabla\varphi$ (once $f$ is expanded as in~\eqref{decomposition}). Differentiable basis functions such as high-order
B-splines ensure that $\nabla\varphi$ is well-defined and continuous, which is 
a key advantage over the use of voxels in geometry optimization.

The expression of $\nabla_\vd[\Proj{f}]$ involves the factor $t$ inside the integral, which depends on the global parameterization of the ray from its intercept $\vo$.
To make it compatible with cell-by-cell ray tracing, we use a re-parameterization that we refer to as the \textit{reference-point translation}.

    \subsection{Reference-Point Translation}
\label{sec:ref_point}
For any $t_0\in\R$, it holds that
\begin{align}
    \nabla_\vd[\Proj{f}](\vo,\vd)
    ={}& (\mathbf{I} - \vd\vd^\top)
    \bigg(
    \int_\R (t-t_0)\nabla f(\vo + t\vd)\dd t \notag \\
    &\mbox{}+ t_0 \int_\R \nabla f(\vo + t\vd)\dd t
    \bigg) \notag \\
    ={}& (\mathbf{I} - \vd\vd^\top)
     \int_\R t \nabla f(\vo + (t+t_0)\vd)\dd t \notag \\
    &\mbox{}+ t_0 \nabla_\vo[\Proj{f}](\vo,\vd) \notag \\[1ex]
    ={}& \nabla_\vd[\Proj{f}](\vo+t_0\vd,\vd) + t_0\nabla_\vo[\Proj{f}](\vo,\vd). \label{eq:second_part}
\end{align}
The second equality uses the fact that $\nabla_\vo[\Proj{f}](\vo, \vd)$ is 
orthogonal to $\vd$, which follows from the fundamental theorem of calculus 
applied to the compactly supported function $f$:
$\int_\R \vd^\top \nabla f(\vo + t\vd)\,\dd t = 0$.

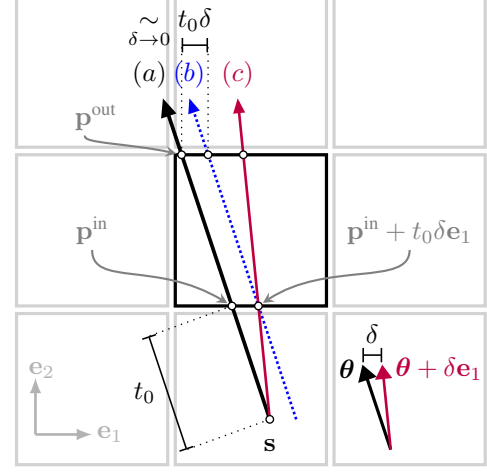
\begin{figure}[t]
    \centering
    \begin{tikzpicture}[scale=0.5]
        
        \foreach \i in {-1, 0, 1} {
            \foreach \j in {-1, 0, 1} {
                \draw[lightgray!70, line width=1.2pt] (\i*4.2, \j*4.2) rectangle (\i*4.2+4, \j*4.2+4);
            }
        }
        \draw[black, line width=1.2pt] (0, 0) rectangle (4, 4);
        
        \coordinate (O) at (2.5, -3);          
        \coordinate (Oblue) at (3.2, -3);      
        
        \coordinate (Pin) at (1.5, 0);         
        \coordinate (Pprimein) at (2.2, 0);    
        
        \coordinate (Pout) at (0.167, 4);      
        \coordinate (Pprimeout) at (1.8, 4);   
        \coordinate (Pblueout) at (0.867, 4);  
        
        \coordinate (Ray1End) at (-0.33, 5.5);
        \coordinate (Ray2End) at (1.65, 5.5);
        \coordinate (RayBlueEnd) at (0.37, 5.5);
        
        \draw[black, line width=1.4pt, -{Triangle[length=2.5mm,width=2mm]}] (O) -- (Ray1End) node[above=0pt, xshift=-5pt] {$(a)$};
        \draw[blue, line width=1.0pt, densely dotted, -{Triangle[length=2mm,width=1.5mm]}] (Oblue) -- (RayBlueEnd) node[above=0pt] {$(b)$};
        \draw[purple, line width=1.0pt, -{Triangle[length=2mm,width=1.5mm]}] (O) -- (Ray2End) node[above=0pt] {$(c)$};
        
        \node[below=4pt] at (O) {$\vo$};
        
        \coordinate (O_shift) at ($(O) + (-2.2, -0.8)$);
        \coordinate (Pin_shift) at ($(Pin) + (-2.2, -0.8)$);
        \draw[line width=0.5pt, dotted] (O) -- (O_shift);
        \draw[line width=0.5pt, dotted] (Pin) -- (Pin_shift);
        \draw[line width=0.6pt, |-|] (O_shift) -- (Pin_shift) node[midway, left=1pt] {$t_0$};
        
        \draw[line width=0.5pt, dotted] (Pout) -- ($(Pout) + (0, 2.9)$);
        \draw[line width=0.5pt, dotted] (Pblueout) -- ($(Pblueout) + (0, 2.9)$);
        \draw[line width=0.6pt, |-|] ($(Pout) + (0, 2.9)$) -- ($(Pblueout) + (0, 2.9)$) 
            node[midway, above, yshift=-4.5pt] {\makebox[0pt][r]{$\underset{\delta \to 0}{\sim}\;$}$t_0 \delta$};
        
        \tikzstyle{dot}=[circle, draw=black, fill=white, inner sep=1pt, line width=0.5pt]
        \node[dot] at (O) {}; \node[dot] at (Pin) {}; \node[dot] at (Pprimein) {};
        \node[dot] at (Pout) {}; \node[dot] at (Pprimeout) {}; \node[dot] at (Pblueout) {};
        
        \node[gray, fill=white, inner sep=2pt] (lbl_pin) at (-2.2, 2.) {$\mathbf{p}^{\text{in}}$};
        \draw[-stealth, gray, thick, shorten >=2pt] (lbl_pin) .. controls (-2.2, 0.5) and (-0.5, 1.2) .. (Pin);

        \node[gray, fill=white, inner sep=2pt] (lbl_pout) at (-2.1, 5.0) {$\mathbf{p}^{\text{out}}$};
        \draw[-stealth, gray, thick, shorten >=2pt] (lbl_pout) .. controls (-2.5, 4.3) and (-1.2, 4.3) .. (Pout);

        \node[gray, fill=white, inner sep=2pt] (lbl_pprimein) at (6.2, 2.0) {$\mathbf{p}^{\text{in}} + t_0 \delta \mathbf{e}_1$};
        \draw[-stealth, gray, thick, shorten >=2pt] (lbl_pprimein) .. controls (6.2, 0.5) and (4.0, 1.2) .. (Pprimein);
        
        \begin{scope}[shift={(5.7, -3.8)}]
            
            \coordinate (T_black)  at (-0.75, 2.25); 
            \coordinate (T_purple) at (-0.225, 2.25); 
            
            \draw[-{Triangle}, black, line width=1.2pt] (0,0) -- (T_black);
            \draw[-{Triangle}, purple, line width=1.0pt] (0,0) -- (T_purple);
            
            \node[black, anchor=east] at (-0.7, 2.15) {$\vd$};
            \node[purple, anchor=west] at (-0.125, 2.15) {$\vd + \delta \mathbf{e}_1$};
            
            \draw[line width=0.6pt, |-|] ($(T_black) + (0, 0.25)$) -- ($(T_purple) + (0, 0.25)$) node[midway, above=1pt] {$\delta$};
                    
        \end{scope}
        \begin{scope}[shift={(-3.7, -3.4)}]
  
            \draw[-latex, gray!60, line width=1.0pt] (0,0) -- (1.5,0) node[right=-2pt, gray!60] {$\mathbf{e}_1$};
            \draw[-latex, gray!60, line width=1.0pt] (0,0) -- (0,1.5);
            \node[gray!60, right, yshift=10pt, xshift=-6pt] at (0, 1.0) {$\mathbf{e}_2$};
        \end{scope}

        \node[fill=white, draw=gray!40, line width=0.8pt, rounded corners=2pt, inner sep=6pt, align=left, anchor=north west] at (-6.5, 12.5) {\footnotesize
             $\color{black}{(a)}= \Proj{f(\vo, \vd)}, 
            \color{blue}{(b)}= \Proj{f(\vo + t_0\delta\mathbf{e}_1, \vd)},
            \color{purple}{(c)}= \Proj{f(\vo, \vd + \delta\mathbf{e}_1)}$
        };
        \node[fill=white, inner sep=6pt, align=left, anchor=north west] at (-6.5, 10.7) {
             $\partial_{\theta_1}[\Proj{f}](\vo, \vd) = \underset{\delta\rightarrow 0}{\lim} \tfrac{\textcolor{purple}{(c)}-\textcolor{black}{(a)}}{\delta}=\underset{\delta\rightarrow 0}{\lim} \tfrac{\textcolor{purple}{(c)}-\textcolor{blue}{(b)}}{\delta} + \underset{\delta\rightarrow 0}{\lim} \tfrac{\textcolor{blue}{(b)}-\textcolor{black}{(a)}}{\delta}$
        };
        
    \end{tikzpicture}
    \caption{Elements involved in the reference-point translation.}
    \label{fig:trick}
\end{figure}

By choosing $t_0$ as the distance from $\vo$ to the entry point of the traversed
cell, we localize the computation to that cell and make the algorithm compatible 
with DDA. This concept is illustrated in Figure~\ref{fig:trick}, and can be expressed formally as
\begin{align}
    \partial_{\theta_1}[\Proj{f}](\vo, \vd) =& \lim_{\delta \to 0}\tfrac{\Proj{f}(\vo, \vd + \delta\mathbf{e}_1) - \Proj{f}(\vo, \vd)}{\delta} \notag\\
    =& \lim_{\delta \to 0}\tfrac{\Proj{f}(\vo, \vd + \delta\mathbf{e}_1) - \Proj{f}(\vo + t_0\delta\mathbf{e}_1, \vd)}{\delta} \notag\\
    &\mbox{}+ \lim_{\delta \to 0}\tfrac{\Proj{f}(\vo + t_0\delta\mathbf{e}_1, \vd)
    - \Proj{f}(\vo, \vd)
    }{\delta} \notag\\
=& \partial_{\theta_1}[\Proj{f}](\underbrace{\vo + t_0\vd}_{=\,\mathbf{p}^{\text{in}}}, \vd) + t_0\partial_{s_1}[\Proj{f}](\vo, \vd). \notag
\end{align}
The direction gradient thus decomposes into a cell-local term and a term that is
proportional to the intercept gradient.

\subsection{Proposed Algorithm}
We derive the full ray-tracing algorithms by combining Section~\ref{sec:ref_point} with the expansion in \eqref{decomposition}. We have that
\begin{align}
    \nabla_\vo[\Proj{f}](\vo, \vd) &= \sum_{\vk\in\Omega} c_\vk \nabla_\vo[\Proj{\varphi}]
    (\vo_{(\vk)}, \vd) \notag\\
    &= \sum_{\vk\in\Omega} c_\vk \bigg(
    \underbrace{
        \int_\R \nabla\varphi(\vo_{(\vk)} + t\vd)\,\dd t
    }_{\texttt{Compute\_grad}_{(\vo)}}
    \bigg), \label{eq:grad_s}
\end{align}
where $\vo_{(\vk)} =\vo - \text{Proj}_{H_{\vd}}(\vk)$. Similarly, it holds that
\begin{align}
    \nabla_\vd&[\Proj{f}](\vo, \vd) = \sum_{\vk\in\Omega} c_\vk \nabla_\vd[\Proj{\varphi}]
    (\vo_{(\vk)}, \vd) \notag\\
    =& (\mathbf{I} - \vd\vd^\top)
    \sum_{\vk\in\Omega} c_\vk \bigg(
    \underbrace{
        \int_\R t\,\nabla\varphi(\vo_{(\vk)} + (t+t_{0,\vk} )\vd)\,\dd t
    }_{\texttt{Compute\_grad}_{(\vd)}} \notag\\
    &\mbox{} + 
    t_{0,\vk} 
    \underbrace{
        \int_\R \nabla\varphi(\vo_{(\vk)} + t\vd)\,\dd t
    }_{\texttt{Compute\_grad}_{(\vo)}}
    \bigg),\label{eq:grad_d}
\end{align}
\begin{figure}[H]
\centering
\begin{tcolorbox}[colback=gray!5, colframe=black!50, fontupper=\small\ttfamily, boxrule=0.4pt, left=4pt, right=4pt, top=3pt, bottom=3pt, width=\columnwidth]
P = project($\mathbf{p}^\text{in}, \vd$) \\
Compute\_grad$_{(\vo)}$ = backward(P, var=$\mathbf{p}^\text{in}$)\\
Compute\_grad$_{(\vd)}$ = backward(P, var=$\vd$)
\end{tcolorbox}
\caption{Code snippet.}
\label{fig:ad_snippet}
\end{figure}

\noindent where $t_{0,\vk} = \|\mathbf{p}^{\text{in}} - \vo_{(\vk)}\|_2$ is the ray parameter at the 
entry point of the cell $\vk$. 
Expressions \eqref{eq:grad_s} and \eqref{eq:grad_d} accumulate cell-wise 
integrals of $\nabla\varphi$ along the ray, making them computable by DDA. 
Algorithms~\ref{alg1} and \ref{alg2} detail these procedures, where 
$\texttt{DDA\_step}(\mathbf{p}^{\text{in}}, \vd)$ returns the exit point 
$\mathbf{p}^{\text{out}}$ of the cell. \label{sec:ad}
The subroutines $\texttt{Compute\_grad}_{(\vo)}$ and $\texttt{Compute\_grad}_{(\vd)}$ are basis-specific and can be obtained in two ways:

\begin{algorithm}[t]
\caption{Compute $\nabla_\vo[\Proj{f}](\vo, \vd)$}
\label{alg1}
\begin{algorithmic}[1]
\Require Volume $c$, intercept $\vo$, direction $\vd$
\State $\mathbf{p}^{\text{in}}=\texttt{entry\_point}(c, \vo, \vd)$ 
\State $\mathbf{g}_\vo = [0,0,0]^\top$
\While{\texttt{DDA} is active}
    \State $\mathbf{p}^{\text{out}} \leftarrow \texttt{DDA\_step}(\mathbf{p}^{\text{in}}, \vd)$
    \State $\mathbf{k} \leftarrow \lfloor (\mathbf{p}^{\text{in}} + \mathbf{p}^{\text{out}})/{2} \rfloor$
        \State $\mathbf{g}_\vo \leftarrow \mathbf{g}_\vo + c[\mathbf{k}]\cdot\texttt{Compute\_grad}_{(\vo)}(\mathbf{k}, \mathbf{p}^{\text{in}}, \mathbf{p}^{\text{out}}, \vd)$

    \State $\mathbf{p}^{\text{in}} \leftarrow \mathbf{p}^{\text{out}}$
\EndWhile
\State \Return $\mathbf{g}_\vo$
\end{algorithmic}
\end{algorithm}

\begin{algorithm}[t]
\caption{Compute $\nabla_\vd[\Proj{f}] (\vo, \vd)$}
\label{alg2}
\begin{algorithmic}[1]
\Require Volume $c$, intercept $\vo$, direction $\vd$
\State $\mathbf{p}^{\text{in}}=\texttt{entry\_point}(c, \vo, \vd)$ 
\State $\mathbf{g}_\vd^{} = [0,0,0]^\top$
\While{\texttt{DDA} is active}
    \State $\mathbf{p}^{\text{out}} \leftarrow \texttt{DDA\_step}(\mathbf{p}^{\text{in}}, \vd)$
    \State $\mathbf{k} \leftarrow \lfloor (\mathbf{p}^{\text{in}} + \mathbf{p}^{\text{out}})/{2} \rfloor$
    \State $t_0 \leftarrow \|\mathbf{p}^{\text{in}} - \vo\|_2$ 
    \State $\mathbf{g}_\vd^{(t_0)} \leftarrow \texttt{Compute\_grad}_{(\vd)}(\mathbf{k}, \mathbf{p}^{\text{in}}, \mathbf{p}^{\text{out}}, \vd)$
    \State $\mathbf{g}_\vo \leftarrow \texttt{Compute\_grad}_{(\vo)}(\mathbf{k}, \mathbf{p}^{\text{in}}, \mathbf{p}^{\text{out}}, \vd)$
    \State $\mathbf{g}_\vd^{} \leftarrow \mathbf{g}_\vd^{} + c[\mathbf{k}] \big(\mathbf{g}_\vd^{(t_0)} + t_0 \mathbf{g}_\vo \big)$
    \State $\mathbf{p}^{\text{in}} \leftarrow \mathbf{p}^{\text{out}}$
\EndWhile
\State \Return $(\mathbf{I} - \vd\vd^\top) \mathbf{g}_\vd^{}$ \Comment{Project onto tangent space}
\end{algorithmic}
\end{algorithm}

\begin{enumerate}
     \item When $\varphi$ admits a closed-form line integral, we implement a function
$\texttt{project}(\mathbf{p}^\text{in}, \vd)$ that computes $ \Proj{\varphi}(\mathbf{p}^\text{in}, \vd)$. 
For spline basis functions, this reduces to the evaluation of appropriate polynomials~\cite{splines_closed_form, horbelt}. The subroutines then follow if one differentiates $\texttt{project}$ with respect to $\mathbf{p}^\text{in}$ and $\vd$, which automatic differentiation (AD) handles at the cost of a single scalar backward pass. This is described in Figure~\ref{fig:ad_snippet}.
    Crucially, AD is applied only to this scalar function, not to the $\texttt{DDA}$ loop itself, so no loop unrolling or graph storage is required. The cost of each subroutine scales linearly with the width of the discrete support of $\varphi$~\cite{haouchattci}.
    \item The subroutines can also be derived analytically when $\nabla\varphi$ is known 
in closed form, as we do for voxels in Section~\ref{sec:voxels}. In this case, there is no need for AD.
\end{enumerate}

\subsection{Closed Form for Voxels}
\label{sec:voxels}
Even though the voxel model is not differentiable in the classic sense, the x-ray transform integrates out the Diracs induced by differentiation. In this case, the volume is discretized with the help of the indicator function \mbox{$\varphi(\cdot) = \mathds{1}_{[0,1]^3}(\cdot)$}. Its gradient expands into a vector of Dirac delta distributions situated on the six faces of the unit cube, with
\begin{equation}
    \nabla\varphi(\vx) = 
    \begin{pmatrix}
        \mathds{1}_{[0,1]^2}(x_2, x_3) \big( \delta(x_1) - \delta(x_1-1) \big) \\
        \mathds{1}_{[0,1]^2}(x_1, x_3) \big( \delta(x_2) - \delta(x_2-1) \big) \\
        \mathds{1}_{[0,1]^2}(x_1, x_2) \big( \delta(x_3) - \delta(x_3-1) \big)
    \end{pmatrix}.
\end{equation}

\begin{algorithm}[t]
\caption{$\texttt{Compute\_grad}_{(\vo)}$ for voxels}
\label{alg:vox1}
\begin{algorithmic}[1]
\Require Index $\mathbf{k}$, entry $\mathbf{p}^{\text{in}}$, exit $\mathbf{p}^{\text{out}}$, direction $\vd$
\State $\boldsymbol{\sigma} = \mathds{1}_{\{\mathbf{p}^{\text{in}} = \mathbf{k}\}} + \mathds{1}_{\{\mathbf{p}^{\text{out}} = \mathbf{k}\}} - \mathds{1}_{\{\mathbf{p}^{\text{in}} = \mathbf{k} + \mathbf{1}\}} - \mathds{1}_{\{\mathbf{p}^{\text{out}} = \mathbf{k} + \mathbf{1}\}}$
\State \Return $\boldsymbol{\sigma} \oslash |\vd|$ \Comment{Element-wise division}
\end{algorithmic}
\end{algorithm}

\begin{algorithm}[t]
\caption{$\texttt{Compute\_grad}_{(\vd)}$ for voxels}
\label{alg:vox2}
\begin{algorithmic}[1]
\Require Index $\mathbf{k}$, entry $\mathbf{p}^{\text{in}}$, exit $\mathbf{p}^{\text{out}}$, direction $\vd$
\State $L = \|\mathbf{p}^{\text{out}} - \mathbf{p}^{\text{in}}\|_2$
\State $\boldsymbol{\sigma}^{\text{out}} = \mathds{1}_{\{\mathbf{p}^{\text{out}} = \mathbf{k}\}} - \mathds{1}_{\{\mathbf{p}^{\text{out}} = \mathbf{k} + \mathbf{1}\}}$
\State \Return $L ( \boldsymbol{\sigma}^{\text{out}} \oslash |\vd|)$ \Comment{Unprojected local gradient}
\end{algorithmic}
\end{algorithm}

Rays that traverse any of the grid planes or intersect the voxel at its edges or corners constitute a set of measure zero; their gradients evaluate to zero and are safely ignored in practice.
Through a change of variables, the integral of $\nabla\varphi$ along a ray parameterized by $\vo$ and $\vd$ evaluates the intersections with the voxel faces as
\begin{equation}
    \int_\R \nabla\varphi(\vo+t\vd)\dd t = 
    \begin{pmatrix}
        \frac{1}{|\theta_1|} (\sigma_\text{left} - \sigma_\text{right})\\
        \frac{1}{|\theta_2|} (\sigma_\text{bottom} - \sigma_\text{top})\\
        \frac{1}{|\theta_3|} (\sigma_\text{back} - \sigma_\text{front})
    \end{pmatrix},
\end{equation}
where the Boolean indicator $\sigma_\text{\emph{face}} \in \{0, 1\}$ is $1$ when the ray intersects the corresponding \emph{face} of the indicator function, and $0$ otherwise. 
Consequently, if the ray either misses the voxel or traverses it by entering and exiting through parallel faces (e.g., entering the left face and exiting the right face), then the components cancel, which leads to 
\begin{equation}
    \int_\R \nabla\varphi(\vo+t\vd)\dd t = \boldsymbol{0}.
\end{equation}
\begin{figure*}[t]
    \centering
    \begin{tikzpicture}

    \tikzset{
        imgnode/.style={draw=black, line width=0.8pt, inner sep=0pt, outer sep=0pt},
        fwd/.style={-{Triangle[length=2.5mm,width=2mm]}, line width=1pt, black},
        grad/.style={-{Triangle[length=2.8mm,width=2.2mm]}, line width=1.6pt, black},
        lbl/.style={font=\small},
        sublbl/.style={font=\footnotesize, black!50},
    }

    \def\xinput{0}
    \def\xmerge{2.0}
    \def\xAop{3.6}
    \def\xsino{5.8}
    \def\xloss{9.0}

    \def\ygeom{1.6}
    \def\yvol{-1.6}
    \def\ygradtop{2.2}
    \def\ygradbot{-2.2}
    \def\ymid{0}
    \def\imgsize{1.8cm}

    \node[imgnode, minimum width=\imgsize, minimum height=\imgsize, fill=white]
        (geombox) at (\xinput, \ygeom) {};

    \begin{scope}
        \clip ($(geombox.south west)+(0.02,0.02)$) rectangle ($(geombox.north east)+(-0.02,-0.02)$);
        \foreach \i in {-3,...,3} {
            \draw[black!20, line width=0.3pt]
                ($(geombox.center)+(\i*0.26, -1)$) -- ($(geombox.center)+(\i*0.26, 1)$);
            \draw[black!20, line width=0.3pt]
                ($(geombox.center)+(-1, \i*0.26)$) -- ($(geombox.center)+(1, \i*0.26)$);
        }
    \end{scope}

    \fill[black] ($(geombox.west)+(-0.6, 0)$) circle (2.5pt);
    \node[font=\scriptsize, anchor=east] at ($(geombox.west)+(-0.75, 0)$) {$\vo_p$};

    \coordinate (src) at ($(geombox.west)+(-0.6, 0)$);
    \coordinate (rayend) at ($(geombox.east)+(0.5, -0.45)$);

    \draw[-{Stealth[length=2.2mm, width=1.6mm, open]}, black, line width=0.9pt]
        (src) -- (rayend);

    \pgfmathsetmacro{\raylength}{0.6}
    \coordinate (dirarrow_start) at (src);
    \coordinate (dirarrow_end) at ($(src)!\raylength cm!(rayend)$);

    \draw[-{Triangle[length=2.2mm, width=1.6mm, open]}, black, line width=1.8pt]
        (dirarrow_start) -- (dirarrow_end);

    \node[font=\scriptsize, black, anchor=north] at ($(dirarrow_start)!0.5!(dirarrow_end)+(0,-0.08)$) {$\vd_p$};

    \node[lbl, anchor=south] at ($(geombox.north)+(0, 0.08)$)
        {Acquisition geometry $\mathbf{\Theta}$};
    \node[sublbl, anchor=north] at ($(geombox.south)+(0, -0.08)$)
        {$\mathbf{\Theta}_p = (\vo_p,\, \vd_p)$};

    \node[imgnode] (vol) at (\xinput, \yvol)
        {\includegraphics[width=\imgsize, height=\imgsize]{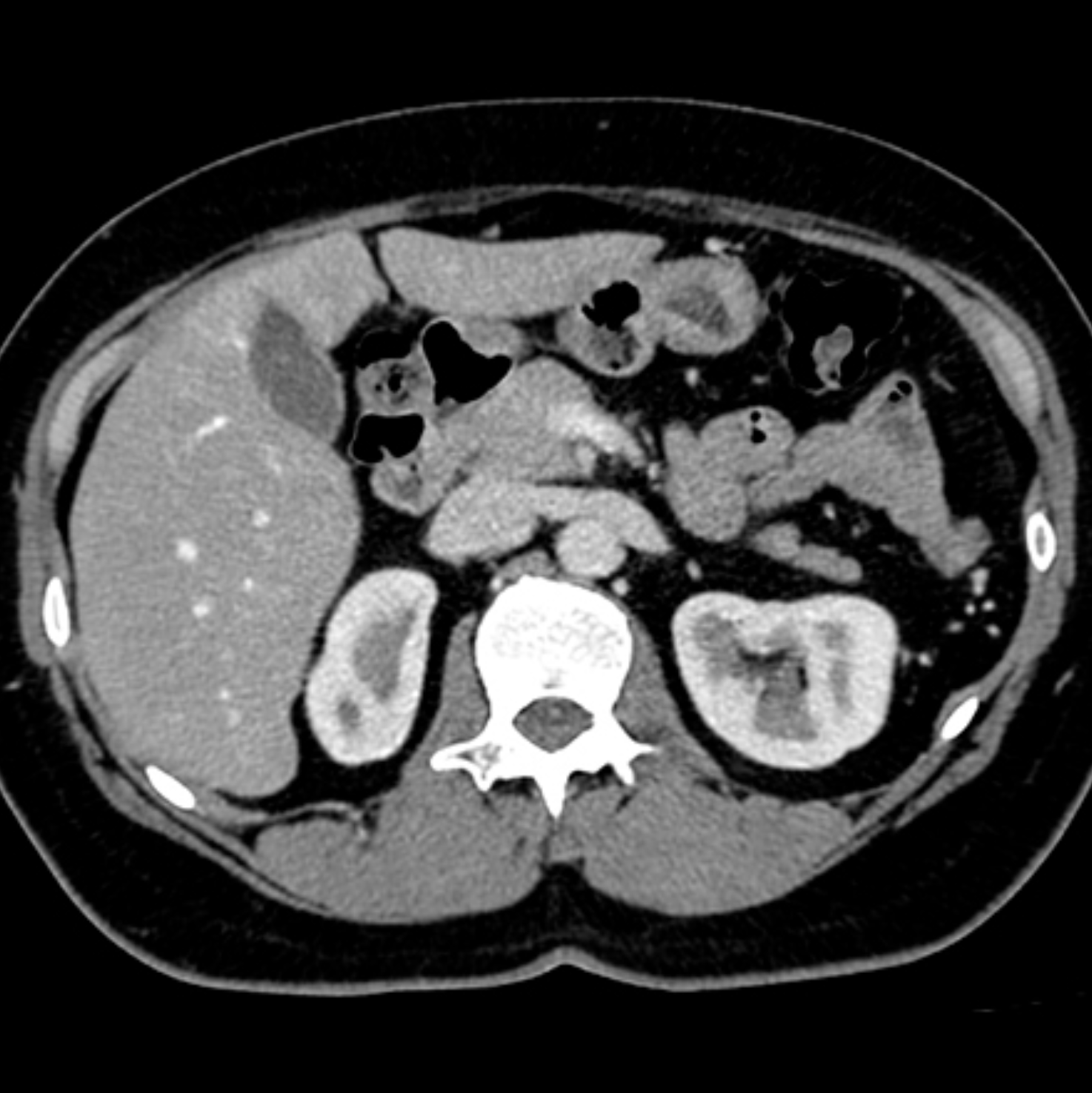}};

    \node[lbl, anchor=south] at ($(vol.north)+(0, 0.08)$)
        {Volume $c$};

    \begin{scope}[shift={(\xinput - 2.7, \yvol-0.8)}, scale=0.28]
        \foreach \i in {0,...,5} {
            \draw[black!30, line width=0.3pt] (\i, 0) -- (\i, 5);
            \draw[black!30, line width=0.3pt] (0, \i) -- (5, \i);
        }
        \def\R{1.5} \def\cx{2.5} \def\cy{2.5}
        \pgfmathsetmacro{\s}{\R*sin(22.5)}
        \pgfmathsetmacro{\c}{\R*cos(22.5)}
        \fill[blue!70, opacity=0.25]
            ({\cx+\c}, {\cy+\s}) -- ({\cx+\s}, {\cy+\c}) -- ({\cx-\s}, {\cy+\c}) -- ({\cx-\c}, {\cy+\s})
            -- ({\cx-\c}, {\cy-\s}) -- ({\cx-\s}, {\cy-\c}) -- ({\cx+\s}, {\cy-\c}) -- ({\cx+\c}, {\cy-\s}) -- cycle;
        \draw[blue!70!black, line width=0.8pt]
            ({\cx+\c}, {\cy+\s}) -- ({\cx+\s}, {\cy+\c}) -- ({\cx-\s}, {\cy+\c}) -- ({\cx-\c}, {\cy+\s})
            -- ({\cx-\c}, {\cy-\s}) -- ({\cx-\s}, {\cy-\c}) -- ({\cx+\s}, {\cy-\c}) -- ({\cx+\c}, {\cy-\s}) -- cycle;
        \fill[blue!70!black] (\cx, \cy) circle (4pt);
        \node[font=\scriptsize, blue!70!black, anchor=south west] at ({\cx+0.01}, {\cy+0.01}) {$\vk$};
        \node[font=\scriptsize, blue!70!black] at (\cx, {\cy-2}) {$c_\vk\,\varphi(\cdot{-}\vk)$};
    \end{scope}

    \node[sublbl, anchor=north] at (\xinput, \yvol-1) {$f = \textstyle\sum_\vk c_\vk\,\varphi(\cdot{-}\vk)$};

    \node[font=\large] (Aop) at (\xAop, \ymid) {$\vA_\mathbf{\Theta}$};
    \node[font=\footnotesize, black!45, anchor=north] at ($(Aop.south)+(0, -0.06)$) {differentiable operator};

    \draw[line width=1pt, black] ($(geombox.east) - (0, 0.7)$) -- (\xmerge, \ygeom-0.7) -- (\xmerge, \ymid);
    \draw[line width=1pt, black] ($(vol.east) + (0, 0.7)$) -- (\xmerge, \yvol+0.7) -- (\xmerge, \ymid);
    \draw[fwd] (\xmerge, \ymid) -- (Aop.west);

    \node[imgnode] (sino) at (\xsino, \ymid)
        {\includegraphics[width=1.1cm, height=\imgsize]{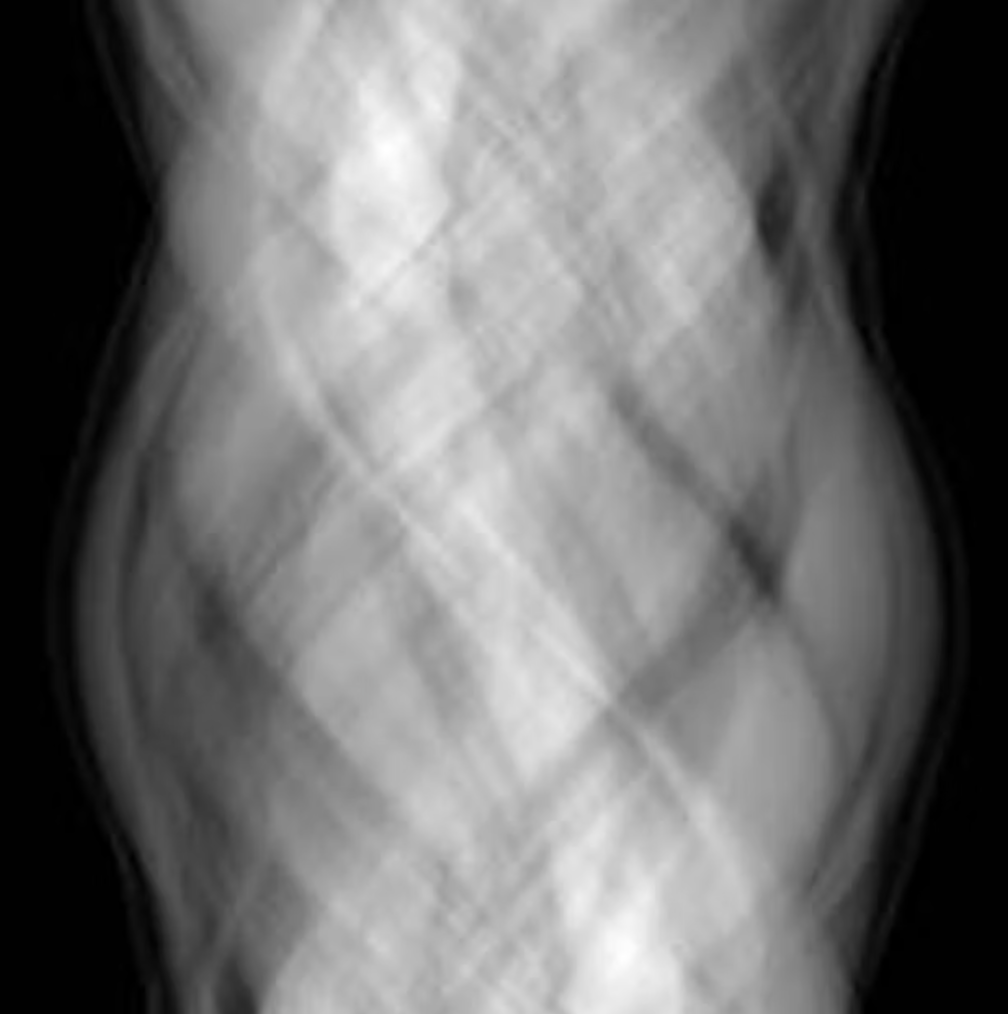}};
    \node[lbl, anchor=south] at ($(sino.north)+(0, 0.08)$) {Projections $\tilde{\vy}$};
    \node[sublbl, anchor=north] at ($(sino.south)+(0, -0.08)$) {$\tilde{\vy} = \vA_\mathbf{\Theta} \vx$};
    \draw[fwd] (Aop.east) -- (sino.west);

    \node[font=\normalsize] (loss) at (\xloss, \ymid) {$\mathcal{L}(\tilde{\vy})$};
    \node[sublbl, anchor=west, align=left] (losseg) at ($(loss.east)+(0.15, 0)$) {$\tfrac{1}{2}\|\tilde{\vy} - \vy\|_2^2$, typically};
    \node[sublbl, anchor=north] at ($(losseg.south)+(0, -0.02)$) {$\vy$: measurements};
    \draw[fwd] (sino.east) -- (loss.west);


    \draw[grad, blue!70!black]
        (loss.north) --
        (loss.north |- {$(geombox.north east) + (0.1, -0.1)$}) --
        ($(geombox.north east) + (0.1, -0.1)$);

    \node[font=\small, fill=white, inner sep=8pt, anchor=south, text=blue!70!black]
        at ({0.5*(\xinput+\xloss)+0.3}, \ygradtop-0.7)
        {$\underbrace{\nabla_\mathbf{\Theta} \vA_\mathbf{\Theta}}_{\text{Alg.~\ref{alg1}\,\&\,\ref{alg2}}}
          \nabla_{\tilde{\vy}} \mathcal{L}$
          \;\; {\small(update $\mathbf{\Theta}$)}};

    \coordinate (targetVol) at ($(vol.south east) + (0.1, +0.1)$);
    \draw[grad]
        (loss.south) --
        (loss.south |- targetVol) --
        (targetVol);

    \node[font=\small, fill=white, inner sep=8pt, anchor=north]
        at ({0.5*(\xinput+\xloss)+0.28}, \ygradbot+0.4)
        {$\underbrace{\vA_\mathbf{\Theta}^\top}_{\text{backprojection}}
        \nabla_{\tilde{\vy}} \mathcal{L}$
          \;\; {\small(update $\vx$)}};

    \end{tikzpicture}
    \caption{Overview of the differentiable framework.}
    \label{fig:framework_overview}
\end{figure*}
Applying the reference-point translation, we compute the angular gradient $\nabla_\vd[\Proj{\varphi}](\vo,\vd)$ by setting the reference intercept $\vo$ to the entry point on the boundary of the voxel of interest. There, the ray parameter is $t = 0$, which causes the entry term to vanish. The integral is then determined exclusively by the exit intersection as 
\begin{equation}
    \int_\R t\nabla\varphi(\vo+(t+t_0)\vd)\dd t = 
    L
    \begin{pmatrix}
        \frac{1}{|\theta_1|} (\sigma_\text{left}^\text{(out)} - \sigma_\text{right}^\text{(out)})\\
        \frac{1}{|\theta_2|} (\sigma_\text{bottom}^\text{(out)} - \sigma_\text{top}^\text{(out)})\\
        \frac{1}{|\theta_3|} (\sigma_\text{back}^\text{(out)} - \sigma_\text{front}^\text{(out)})
    \end{pmatrix},
\end{equation}
where $L$ is the total intersection length between the ray and the voxel, and where $\sigma_\text{\emph{face}}^\text{(out)} \in \{0, 1\}$ equals $1$ if the \emph{exit face} of the ray corresponds to that specific boundary of the voxel, and $0$ otherwise. The corresponding subroutines are given in Algorithms~\ref{alg:vox1} and~\ref{alg:vox2}.

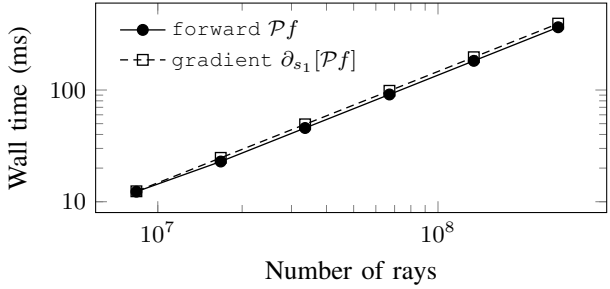
\begin{figure}[t]
\centering
\begin{tikzpicture}
\begin{axis}[
  xmode=log, ymode=log,
  width=0.46\textwidth, height=0.24\textwidth,
  xlabel={Number of rays},
  ylabel={Wall time (ms)},
  xmin=6e6, xmax=4e8,
  ymin=8,   ymax=600,
  xtick={1e7, 1e8},
  xticklabels={$10^7$, $10^8$},
  ytick={10, 100},
  yticklabels={$10$, $100$},
  legend pos=north west,
  legend style={font=\footnotesize, draw=none, fill=none},
  legend cell align=left,
  tick label style={font=\small},
]
\addplot[black, mark=*, line width=0.5pt] coordinates {
    (8388608,    12.3)
    (16777216,   22.9)
    (33554432,   45.7)
    (67108864,   91.2)
    (134217728, 182.4)
    (268435456, 364.6)
  };
\addlegendentry{\texttt{forward} $\Proj{f}$}
\addplot[black, densely dashed, mark=square,mark options={solid}, mark size=2pt, line width=0.5pt] coordinates {
    (8388608,    12.4)
    (16777216,   24.7)
    (33554432,   49.3)
    (67108864,   98.4)
    (134217728, 196.6)
    (268435456, 393.1)
  };
\addlegendentry{\texttt{gradient} $\partial_{s_1}[\Proj{f}]$}
\end{axis}
\end{tikzpicture}
\caption{Computational complexity of a gradient entry with our method.}
\label{fig:speed}
\end{figure}

\subsection{Implementation Details}
\label{sec:implementation}
We implement our ray-tracing operators in Dr.Jit~\cite{drjit}, a just-in-time compiler
that traces computational graphs and compiles efficient CPU and GPU codes. Each ray is processed
independently: an outer loop over rays runs in parallel across threads while
an inner DDA loop runs sequentially within each thread.
Backward-mode automatic differentiation, as in Dr.Jit, PyTorch, or JAX, records every operation
as a directed acyclic graph before it is traversed in reverse. The forward x-ray DDA loop
violates this: its variables (ray--cell intersection points) are updated \textit{in place} at each
step for efficiency. This creates an unfortunate cyclic dependency between the accumulated quantities and the
differentiated variables. 
In this work, we bypass this state of affairs by defining custom operators (gradients with
respect to~$\vo$ and $\vd$) with an explicit vector-Jacobian product (VJP). The VJP itself
(Algorithms~\ref{alg1} and \ref{alg2}) is a ray-tracing pass. The AD framework applies the
chain rule and calls our custom VJP whenever the x-ray transform appears. A general overview
of the framework is shown in Figure~\ref{fig:framework_overview}. Runtime and memory comparisons against full automatic differentiation appear in the Appendix \ref{sec:appendix}.

To demonstrate our claims on complexity, we benchmark our implementation on 3D setups with arbitrary geometry in Fig.~\ref{fig:speed}. Each geometric gradient entry
matches the forward runtime with a mean ratio of $1.07\pm0.03$, confirming that closed-form
differentiation adds no asymptotic overhead.

\section{Numerical Experiments}

Tomographic reconstruction is formulated as the search for an estimate of $\vx\in\R^N$ from measurements $\vy\in\R^M$. The measurements are assumed to follow the model
\begin{equation}
\label{ip}
    \vy = \mathbf{A}_\mathbf{\Theta}\vx + \vn, \quad \vn\sim\mathcal{N}(\mathbf{0}, \sigma^2\mathbf{I}_M),
\end{equation}
where $\vx$ is a vector where components collect the expansion coefficients $c_\vk$ describing the continuous image $f$ as in \eqref{decomposition}, and $\vA_\mathbf{\Theta}\in\R^{M\times N}$ is the x-ray forward operator such that 
\begin{equation}
[\vA_\mathbf{\Theta}]_{m,n} = \Proj{\varphi}(\vo_m - \text{Proj}_{H_{\vd_m}}(\vk_n), \vd_m).
\end{equation}
The linear operator $\mathbf{A}$ is parameterized by the acquisition parameters $\mathbf{\Theta} = \{ (\vo_m, \vd_m)\in \R^{3\times 2}, m=1\ldots M\}$.
Each measurement component $y_m$ can be written as 
\begin{equation}
    y_m =  \Proj{f}(\vo_m, \vd_m) + n_m, \quad m=1, \dotsc, M.
\end{equation}

In practice, a \textit{nominal} geometry is known, which differs from $\mathbf{\Theta}$ by a small perturbation denoted $g$ and expressed as
\begin{equation}
    \mathbf{\Theta}^{\text{nom}} = g(\mathbf{\Theta}). \label{nominal}
\end{equation}
Usually, the acquisition geometry is also \emph{structured}, in the sense that 
the geometric perturbations can be parameterized by fewer variables than the total number of rays.

The uncalibrated solution to the reconstruction problem is typically expressed through a maximum-likelihood formulation as
\begin{equation}
    \vx^\ast = \underset{\vx\in\R^N}{\argmin} \underbrace{\tfrac{1}{2\sigma^2}\|\vy - \vA_\mathbf{\Theta}^{\text{nom}}\vx\|_2^2}_{\mathcal{L}(\vx, \mathbf{\Theta}^{\text{nom}})}.
\end{equation}

Suppose we want at the same time to calibrate the model and to reconstruct the volume. Then, the optimization methods will need to find an estimate of both $\vx$ and $\mathbf{\Theta}$, which will involve maximum-likelihood formulations such as 
\begin{equation}
\label{eq:joint}
    \vx^\ast, \boldsymbol{\Theta}^\ast = \underset{\vx\in\R^N, \boldsymbol{\Theta} \in \R^{M \times 2\times 3}}{\argmin} \mathcal{L}(\vx, \mathbf{\Theta}).
\end{equation}

We conduct calibration experiments in 
both real and simulated settings. 
We solve the non-convex joint problem~\eqref{eq:joint} with Adam~\cite{adam} throughout. All geometric perturbations are expressed in units of a single detector element.

\subsection{Simulated Data}
\begin{figure}[t]
    \centering
    \begin{tikzpicture}
    
        \tikzset{
            lbl/.style={font=\scriptsize}
        }

        \pgfplotsset{
            recon/.style={
                width=0.30\columnwidth, height=0.30\columnwidth,
                scale only axis, enlargelimits=false, axis lines=none,
            }
        }

        \node[lbl] at (0.15\columnwidth, 0.32\columnwidth) {$\sigma_{\text{j}} = 1$};
        \node[lbl] at (0.47\columnwidth, 0.32\columnwidth) {$\sigma_{\text{j}} = 5$};
        \node[lbl] at (0.79\columnwidth, 0.32\columnwidth) {$\sigma_{\text{j}} = 9$};

        
        \node[lbl, rotate=90, anchor=south] at (-0.02\columnwidth, 0.15\columnwidth) {Uncalibrated};

        \begin{axis}[name=n1, recon]
            \addplot graphics[xmin=0, xmax=1, ymin=0, ymax=1] {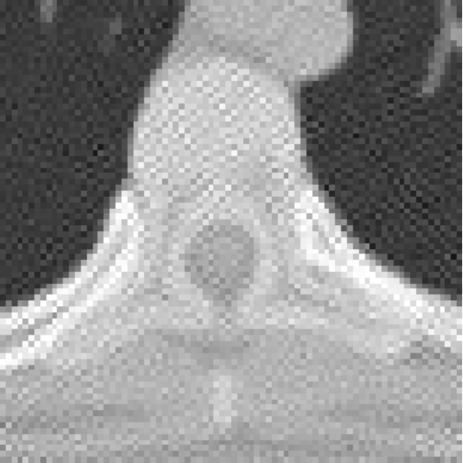};
        \end{axis}
        
        \begin{axis}[name=n5, at={(n1.outer east)}, anchor=outer west, xshift=1mm, recon]
            \addplot graphics[xmin=0, xmax=1, ymin=0, ymax=1] {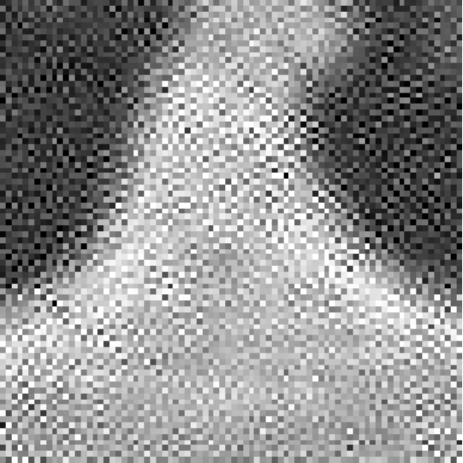};
        \end{axis}
        
        \begin{axis}[name=n10, at={(n5.outer east)}, anchor=outer west, xshift=1mm, recon]
            \addplot graphics[xmin=0, xmax=1, ymin=0, ymax=1] {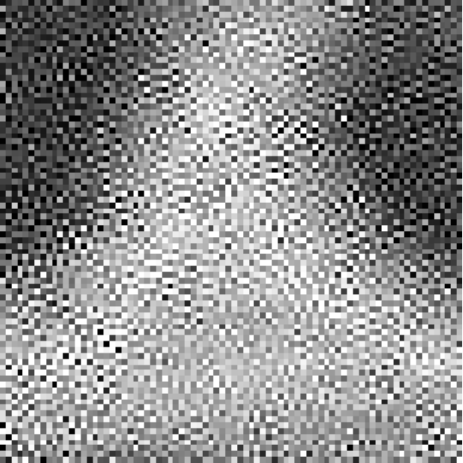};
        \end{axis}

        
        \node[lbl, rotate=90, anchor=south] at (-0.02\columnwidth, -0.16\columnwidth) {Our method};

        \begin{axis}[name=c1, at={(n1.south west)}, anchor=north west, yshift=-1.5mm, recon]
            \addplot graphics[xmin=0, xmax=1, ymin=0, ymax=1] {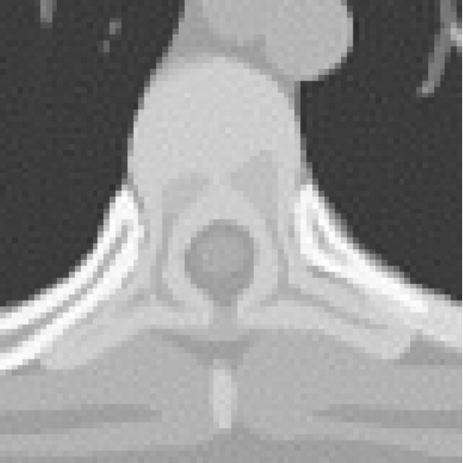};
        \end{axis}
        
        \begin{axis}[name=c5, at={(c1.outer east)}, anchor=outer west, xshift=1mm, recon]
            \addplot graphics[xmin=0, xmax=1, ymin=0, ymax=1] {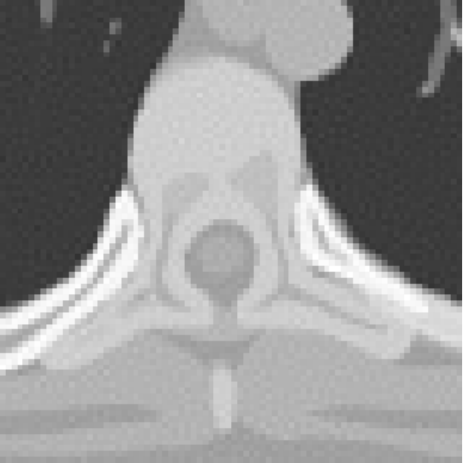};
        \end{axis}
        
        \begin{axis}[name=c10, at={(c5.outer east)}, anchor=outer west, xshift=1mm, recon]
            \addplot graphics[xmin=0, xmax=1, ymin=0, ymax=1] {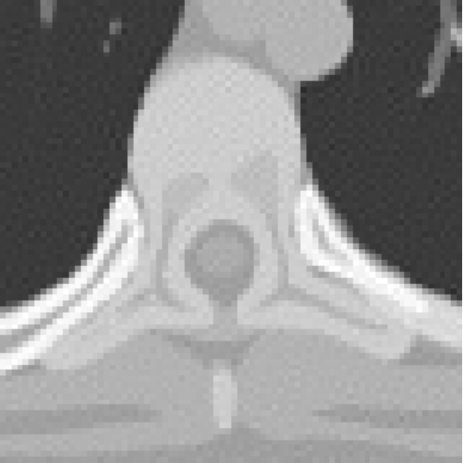};
        \end{axis}
        
    \end{tikzpicture}
    \caption{Synthetic calibration experiment. Top row: reconstructions with the nominal geometry. Bottom row: after joint calibration with our method and quadratic B-splines. Columns correspond to increasing jitter levels \unboldmath$\sigma_{\text{j}} \in \{1, 5, 9\}$ applied to projection angles and detector offsets.}
    \label{fig:cal_simu}
\end{figure}

\begin{figure}[t]
\centering
\begin{tikzpicture}
\pgfplotsset{
    metrics/.style={
        width=0.42\columnwidth,
        height=2.4cm,
        scale only axis,
        grid=major,
        grid style={gray!20},
        xlabel={$\sigma_\text{j}$},
        xtick={1,2,...,9},
        label style={font=\scriptsize},
        tick label style={font=\tiny},
        title style={font=\scriptsize, yshift=-4pt},
    }
}

\begin{axis}[
    metrics,
    name=psnr,
    title={PSNR (dB)},
    ymin=0, ymax=42,
]
\addplot[black!40, thick, dashed, mark=o, mark size=1.2pt] coordinates {
    (1,21.77)(2,17.23)(3,14.54)(4,11.40)(5,11.19)
    (6,9.99)(7,8.98)(8,8.11)(9,7.37)
};
\addplot[black, thick, solid, mark=o, mark size=1.2pt] coordinates {
    (1,37.74)(2,32.90)(3,31.30)(4,29.94)(5,28.62)
    (6,29.04)(7,29.09)(8,28.04)(9,29.03)
};
\end{axis}

\begin{axis}[
    metrics,
    name=ssim,
    at={(psnr.outer east)},
    anchor=outer west,
    xshift=1mm,
    title={SSIM},
    ymin=0, ymax=1.05,
    ytick={0,0.2,0.4,0.6,0.8,1.0},
]
\addplot[black!40, thick, dashed, mark=o, mark size=1.2pt] coordinates {
    (1,0.2729)(2,0.1375)(3,0.0835)(4,0.0502)(5,0.0394)
    (6,0.0293)(7,0.0227)(8,0.0181)(9,0.0147)
};
\addplot[black, thick, solid, mark=o, mark size=1.2pt] coordinates {
    (1,0.9220)(2,0.8704)(3,0.8495)(4,0.8436)(5,0.8281)
    (6,0.8024)(7,0.7989)(8,0.7967)(9,0.7964)
};
\end{axis}

\coordinate (midtop) at ($(psnr.north east)!0.5!(ssim.north west)$);
\node[anchor=south, font=\scriptsize, inner sep=2pt] at (midtop) {%
    \begin{tabular}{l}
        \raisebox{1pt}{\tikz{\draw[black!40,thick,dashed](0,0)--(0.3,0);}} Uncalib.\\[1pt]
        \raisebox{1pt}{\tikz{\draw[black,thick,solid](0,0)--(0.3,0);}} Ours
    \end{tabular}
};

\end{tikzpicture}
\caption{Quality metrics for the simulated CT experiment of Figure~\ref{fig:cal_simu}.}
\label{fig:metrics}
\end{figure}
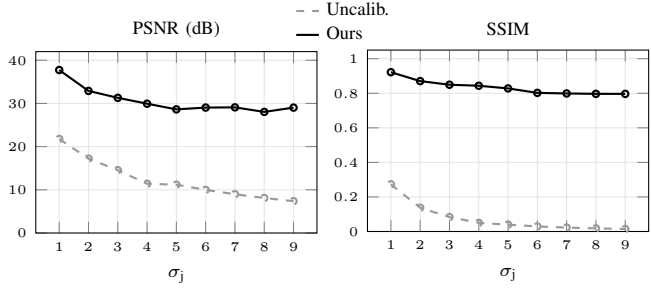

\begin{figure}[t]
    \centering
    \begin{tikzpicture}
    
        \tikzset{
            lbl/.style={font=\scriptsize},
            roi/.style={draw=yellow, line width=0.6pt, dashed}
        }

        \pgfplotsset{
            recon/.style={
                width=0.30\columnwidth, height=0.30\columnwidth,
                scale only axis, enlargelimits=false, axis lines=none,
            },
            zoom/.style={
                width=0.30\columnwidth, height=0.30\columnwidth,
                scale only axis, enlargelimits=false, axis lines=none,
                axis line style={draw=black!30, line width=0.5pt},
                axis lines=box 
            }
        }

        \def\zxmin{0.3} \def\zxmax{0.7}
        \def\zymin{0.1} \def\zymax{0.5}

        \node[lbl] at (0.15\columnwidth, 0.33\columnwidth) {Ground truth};
        \node[lbl] at (0.47\columnwidth, 0.33\columnwidth) {Uncalibrated};
        \node[lbl] at (0.77\columnwidth, 0.33\columnwidth) {Our method};

        
        \begin{axis}[name=n1, recon]
            \addplot graphics[xmin=0, xmax=1, ymin=0, ymax=1] {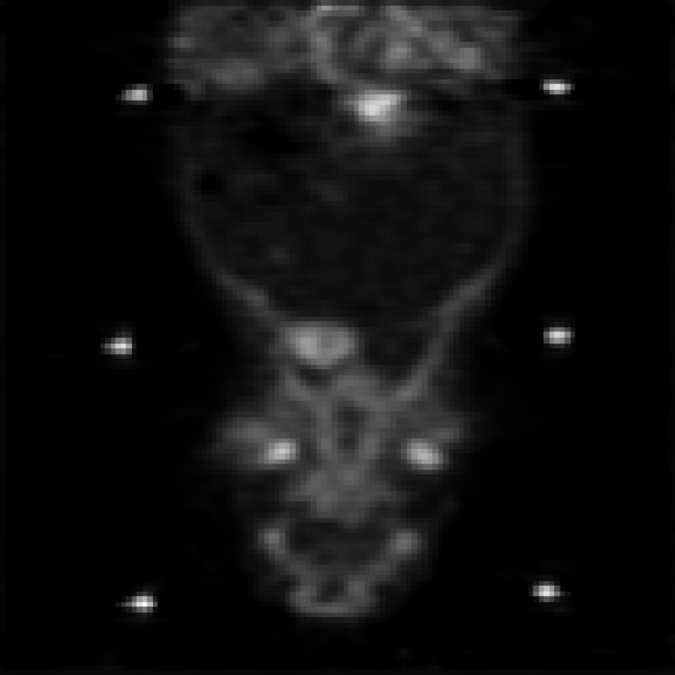};
            \draw[roi] (axis cs:\zxmin,\zymin) rectangle (axis cs:\zxmax,\zymax);
        \end{axis}
        
        \begin{axis}[name=n5, at={(n1.outer east)}, anchor=outer west, xshift=1mm, recon]
            \addplot graphics[xmin=0, xmax=1, ymin=0, ymax=1] {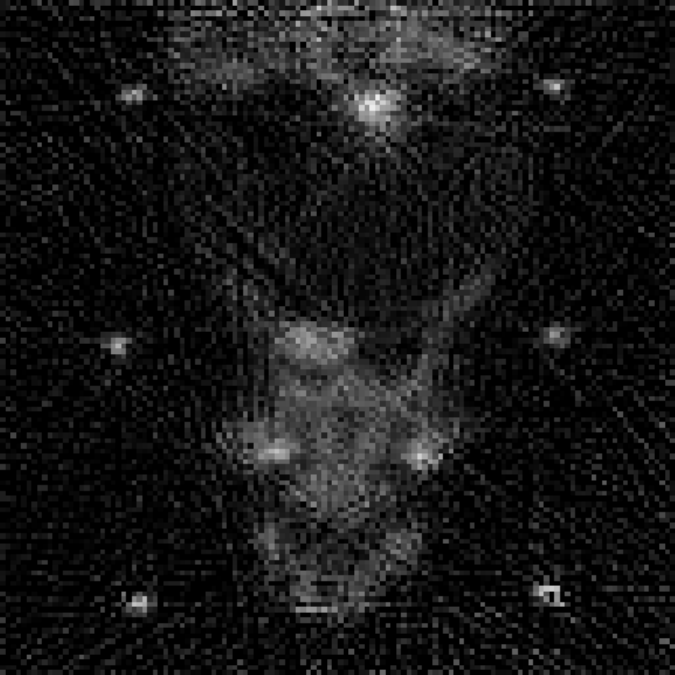};
            \draw[roi] (axis cs:\zxmin,\zymin) rectangle (axis cs:\zxmax,\zymax);
        \end{axis}
        
        \begin{axis}[name=n10, at={(n5.outer east)}, anchor=outer west, xshift=1mm, recon]
            \addplot graphics[xmin=0, xmax=1, ymin=0, ymax=1] {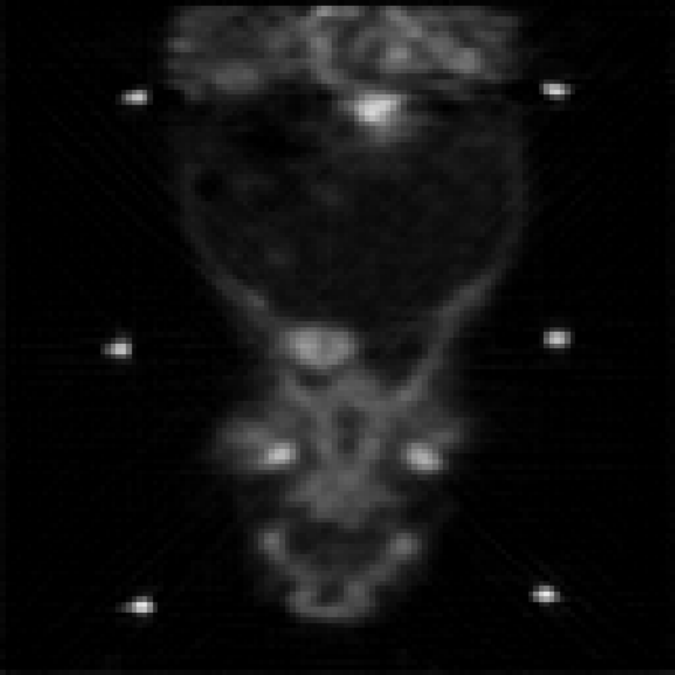};
            \draw[roi] (axis cs:\zxmin,\zymin) rectangle (axis cs:\zxmax,\zymax);
        \end{axis}

        
        \begin{axis}[name=z1, at={(n1.below south)}, anchor=above north, yshift=-2mm, zoom, 
                     xmin=\zxmin, xmax=\zxmax, ymin=\zymin, ymax=\zymax]
            \addplot graphics[xmin=0, xmax=1, ymin=0, ymax=1] {Figures/x_true.pdf};
        \end{axis}

        \begin{axis}[name=z5, at={(n5.below south)}, anchor=above north, yshift=-2mm, zoom,
                     xmin=\zxmin, xmax=\zxmax, ymin=\zymin, ymax=\zymax]
            \addplot graphics[xmin=0, xmax=1, ymin=0, ymax=1] {Figures/x_no_calib.pdf};
        \end{axis}

        \begin{axis}[name=z10, at={(n10.below south)}, anchor=above north, yshift=-2mm, zoom,
                     xmin=\zxmin, xmax=\zxmax, ymin=\zymin, ymax=\zymax]
            \addplot graphics[xmin=0, xmax=1, ymin=0, ymax=1] {Figures/x_with_calib.pdf};
        \end{axis}
        
    \end{tikzpicture}
    \caption{Calibration with PET geometry. Top row: full reconstruction. Bottom row: zoom on the central region. Left: ground truth; center: reconstruction with the perturbed geometry (\unboldmath$\mathcal{L} = 0.8$); right: reconstruction after calibration (\unboldmath$\mathcal{L} = 1.5 \times 10^{-4}$).}
    \label{fig:cal_pet}
\end{figure}

\paragraph{Cone-Beam CT}
We assume a cone-beam geometry while reconstructing a volume of size $512^3$ from $1024$ projection angles on a 
$(768 \times 768)$ detector. The unknown perturbation $g$ 
in~\eqref{nominal} adds independent Gaussian noise of standard deviation 
$\sigma_\text{j} \in \{1, \ldots, 9\}$ to each angle and detector 
offset, as in column~(c) of Figure~\ref{fig:calibration_scenarios}. The insets of
Figure~\ref{fig:cal_simu} contain a region of the uncalibrated and 
calibrated reconstructions for $\sigma_\text{j} \in \{1, 5, 9\}$, obtained 
with quadratic B-splines as basis functions. We report in Figure~\ref{fig:metrics} 
the PSNRs and SSIMs as a function of $\sigma_\text{j}$ for all 
levels.

\paragraph{PET-Like Geometry}
We apply our framework to a PET-like geometry using the DigiMouse 
phantom~\cite{digimouse} discretized on a $128^3$ grid. The detector ring 
consists of $8$ blocks of $(64 \times 64)$ scintillator elements, placed at a 
radius of $64$ pixels from the center. Each measurement is a line of response 
between two detector elements. The perturbation $g$ in~\eqref{nominal} applies 
an independent random displacement of standard deviation $\sigma_\text{j} = 2$ 
to each block, as in column~(d) of Figure~\ref{fig:calibration_scenarios}. 
Reconstructions with and without calibration are shown in Figure~\ref{fig:cal_pet} .

\paragraph{Benefit of Higher-Order Basis Functions}
We compare voxel and quadratic B-spline representations on a parallel-beam 
calibration task using a phantom ($N = 500^3$, $900$ projection angles, 
$(900 \times 900)$ detector) with Gaussian offset jitter level $\sigma_\text{j}=5$. 

\begin{figure}[H]
    \centering
    \begin{tikzpicture}

        \pgfplotsset{
            img/.style={
                width=0.29\columnwidth, height=0.29\columnwidth,
                scale only axis, enlargelimits=false, axis lines=none,
                title style={font=\scriptsize},
            },
            zoom/.style={
                width=0.29\columnwidth, height=0.18\columnwidth,
                scale only axis, enlargelimits=false,
                xtick=\empty, ytick=\empty,
                axis line style={yellow, line width=0.8pt},
                xmin=0.62, xmax=0.78,
                ymin=0.65, ymax=0.75,
            },
        }


        \begin{axis}[img, name=img1, title={Uncalibrated}]
            \addplot graphics[xmin=0,xmax=1,ymin=0,ymax=1]{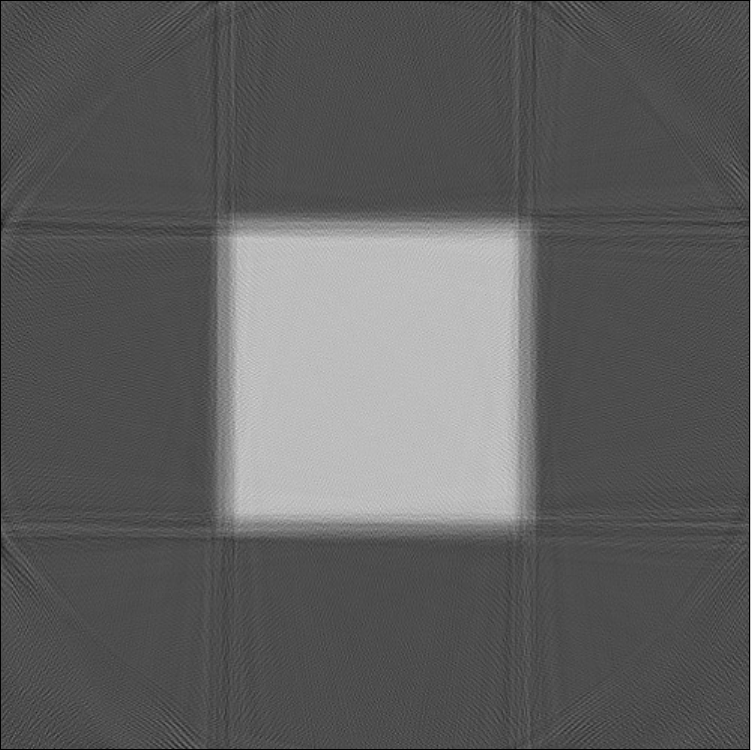};
            \draw[yellow] (0.62,0.65) rectangle (0.78,0.75);
        \end{axis}

        \begin{axis}[img, name=img2,
                     at={(img1.outer east)}, anchor=outer west, xshift=1mm,
                     title={Our method (pixels)}]
            \addplot graphics[xmin=0,xmax=1,ymin=0,ymax=1]{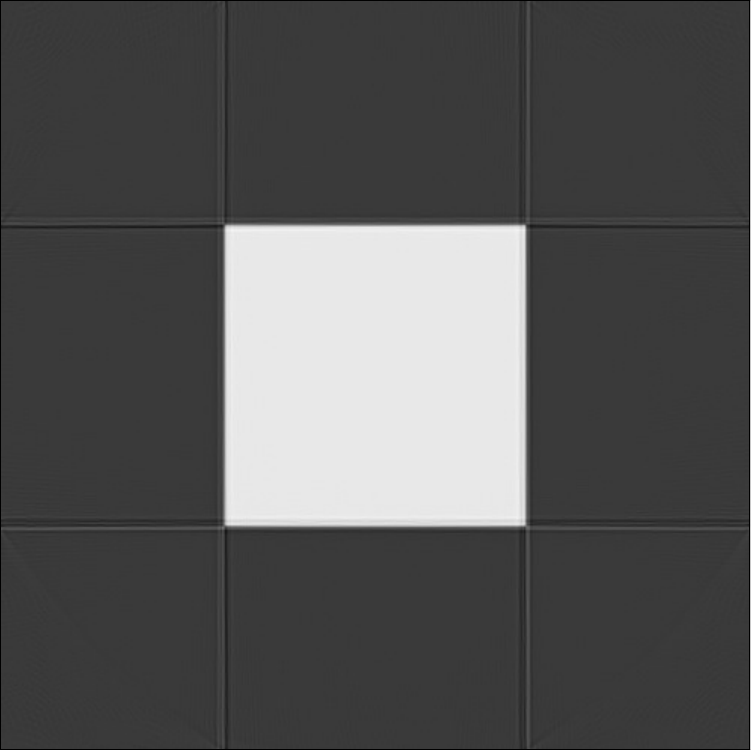};
            \draw[yellow] (0.62,0.65) rectangle (0.78,0.75);
        \end{axis}

        \begin{axis}[img, name=img3,
                     at={(img2.outer east)}, anchor=outer west, xshift=1mm,
                     title={Our method (B-splines)}]
            \addplot graphics[xmin=0,xmax=1,ymin=0,ymax=1]{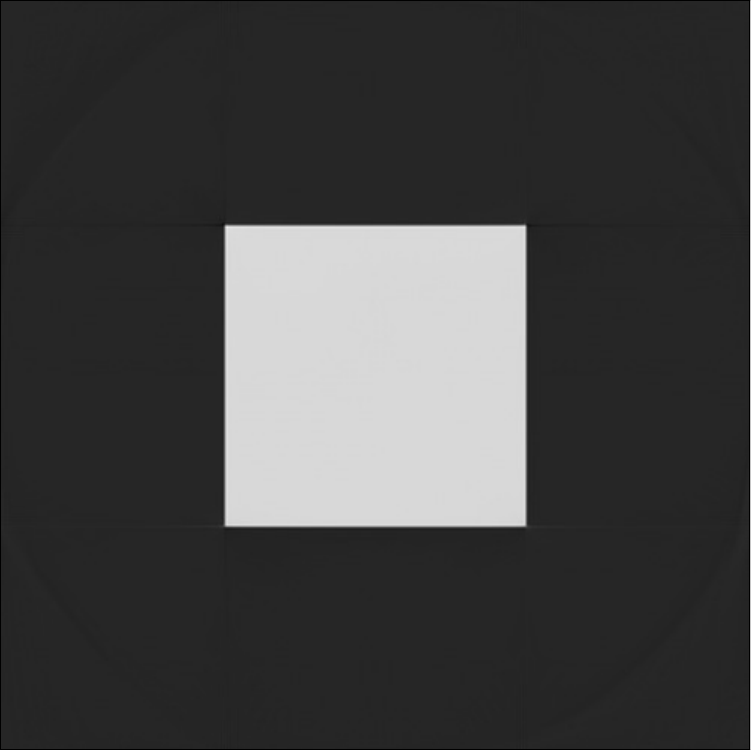};
            \draw[yellow] (0.62,0.65) rectangle (0.78,0.75);
        \end{axis}


        \begin{axis}[zoom, name=zoom1,
                     at={(img1.south west)}, anchor=north west, yshift=-1.5mm]
            \addplot graphics[xmin=0,xmax=1,ymin=0,ymax=1]{Figures/recon_naive.pdf};
        \end{axis}

        \begin{axis}[zoom, name=zoom2,
                     at={(img2.south west)}, anchor=north west, yshift=-1.5mm]
            \addplot graphics[xmin=0,xmax=1,ymin=0,ymax=1]{Figures/recon_calibrated_pixels.pdf};
        \end{axis}

        \begin{axis}[zoom, name=zoom3,
                     at={(img3.south west)}, anchor=north west, yshift=-1.5mm]
            \addplot graphics[xmin=0,xmax=1,ymin=0,ymax=1]{Figures/recon_calibrated_splines2.pdf};
        \end{axis}


        \begin{axis}[
            name=ax1,
            at={(zoom2.south)}, anchor=north,
            yshift=-0.9cm,
            width=0.67\columnwidth, height=3.2cm,
            scale only axis, grid=major,
            xlabel={Shift error (pixels)},
            ylabel={Loss landscape},
            label style={font=\scriptsize},
            tick label style={font=\tiny},
            legend columns=2,
            legend style={
                font=\small,
                cells={anchor=west},
                nodes={scale=0.75, transform shape},
                at={(0.5, 0.97)}, anchor=north,
                fill=white, fill opacity=0.8,
                draw opacity=1, text opacity=1,
                /tikz/every even column/.append style={column sep=0.6em},
            },
        ]
            \addplot[black, thick, solid]
    table[col sep=comma, x={Shift Error (pixels)}, y={Loss Order 0}]
    {Figures/calibration_loss_and_gradients.csv};
\addlegendentry{Pixels}

\addplot[black, thick, solid, mark=o, mark repeat=10, mark size=1.2pt]
    table[col sep=comma, x={Shift Error (pixels)}, y={Loss Order 2}]
    {Figures/calibration_loss_and_gradients.csv};
\addlegendentry{B-splines}
        \end{axis}

    \end{tikzpicture}
    \caption{Benefit of higher-order basis functions for calibration. Top row: 
reconstructions without calibration (left), with pixel-based calibration 
(center), and with quadratic B-spline calibration (right). The insets show 
zoomed regions. Bottom: loss landscape, as a function of a 1D detector shift at angle zero, 
for pixels and B-splines. }
    \label{fig:cal_splines}
\end{figure}
\newlength{\panelH}\setlength{\panelH}{3.8cm}

\begin{figure*}[t]
    \centering

    \begin{minipage}{0.03\linewidth}
        \hfill
    \end{minipage}%
    \hfill
    \begin{minipage}{0.22\linewidth}
        \centering
        {\shortstack{Misaligned COR}}
    \end{minipage}%
    \hfill
    \begin{minipage}{0.22\linewidth}
        \centering
        {\shortstack{Source jitter}}
    \end{minipage}%
    \hfill
    \begin{minipage}{0.22\linewidth}
        \centering
        {\shortstack{Angle imprecision}}
    \end{minipage}%
    \hfill
    \begin{minipage}{0.22\linewidth}
        \centering
        {\shortstack{Detector vibrations}}
    \end{minipage}%

    \vspace{0.5em} 

    \begin{minipage}[c]{0.03\linewidth}
        \centering
        \rotatebox{90}{{Uncalibrated}}
    \end{minipage}%
    \hfill
    \begin{minipage}[c]{0.22\linewidth}
        \centering
        \includegraphics[height=\panelH]{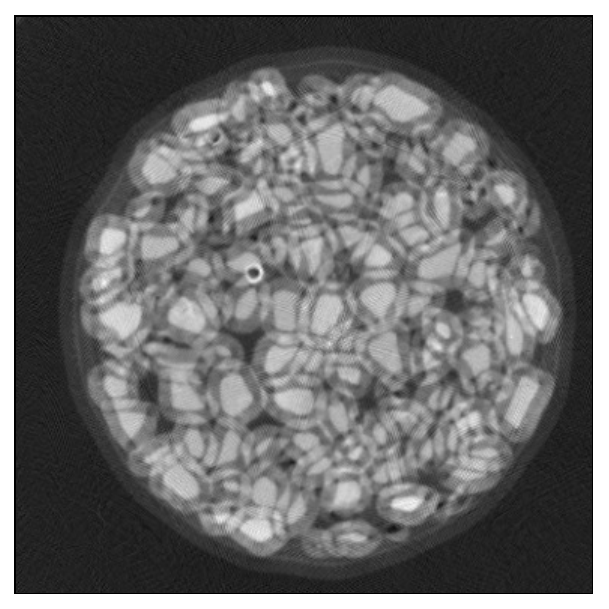}
    \end{minipage}%
    \hfill
    \begin{minipage}[c]{0.22\linewidth}
        \centering
        \includegraphics[height=\panelH]{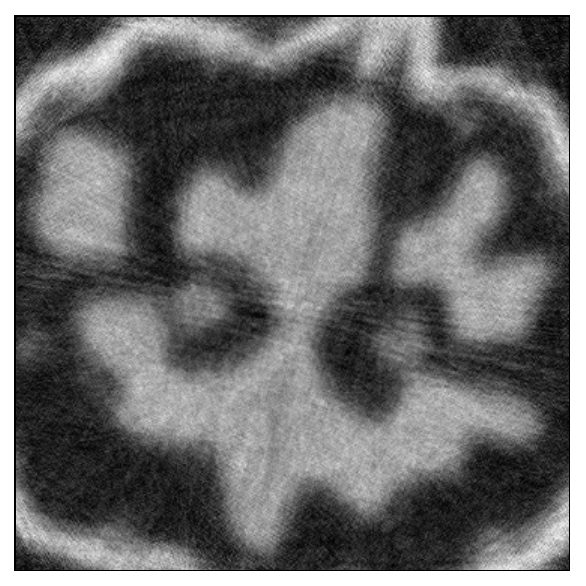}
    \end{minipage}%
    \hfill
    \begin{minipage}[c]{0.22\linewidth}
        \centering
        \includegraphics[height=\panelH]{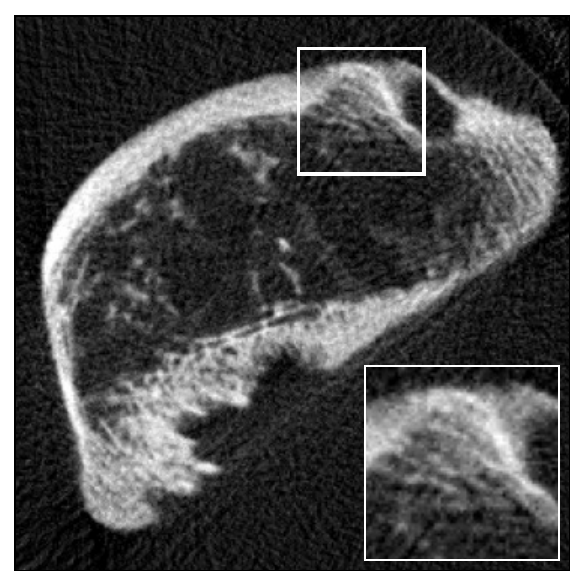}
    \end{minipage}%
    \hfill
    \begin{minipage}[c]{0.22\linewidth}
        \centering
        \includegraphics[height=\panelH]{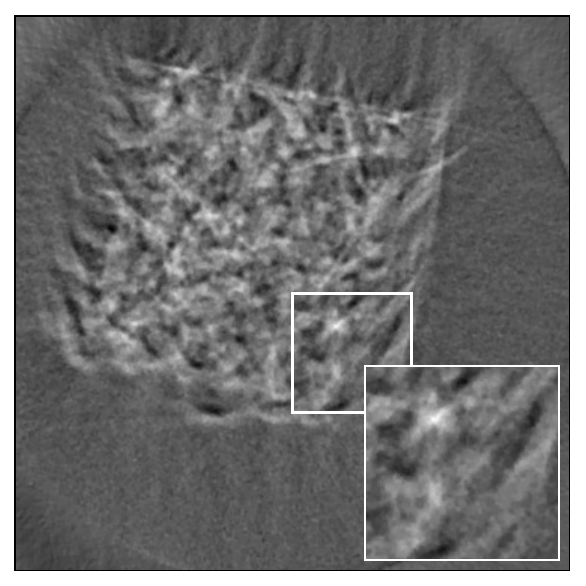}
    \end{minipage}

    \vspace{1em} 

    \begin{minipage}[c]{0.03\linewidth}
        \centering
        \rotatebox{90}{{Our method}}
    \end{minipage}%
    \hfill
    \begin{minipage}[c]{0.22\linewidth}
        \centering
        \includegraphics[height=\panelH]{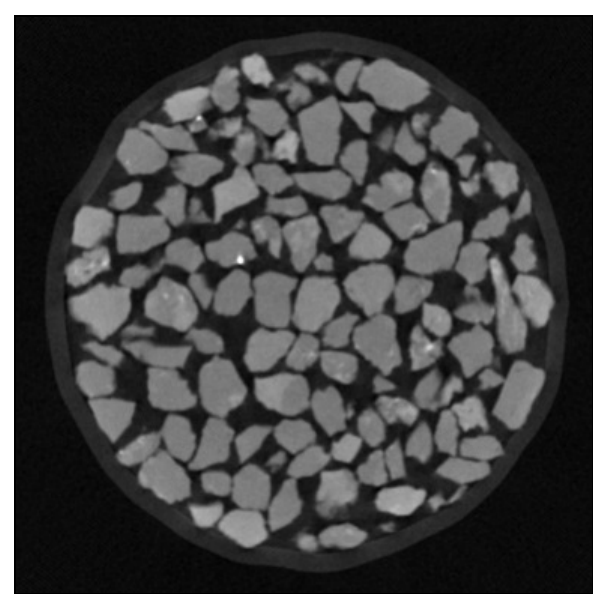}
    \end{minipage}%
    \hfill
    \begin{minipage}[c]{0.22\linewidth}
        \centering
        \includegraphics[height=\panelH]{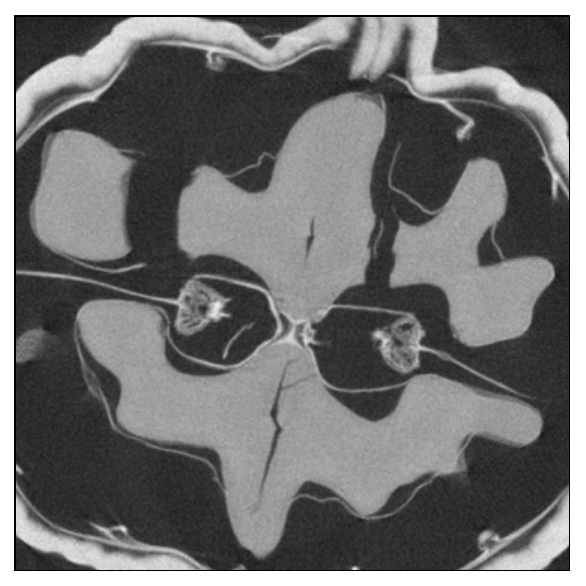}
    \end{minipage}%
    \hfill
    \begin{minipage}[c]{0.22\linewidth}
        \centering
        \includegraphics[height=\panelH]{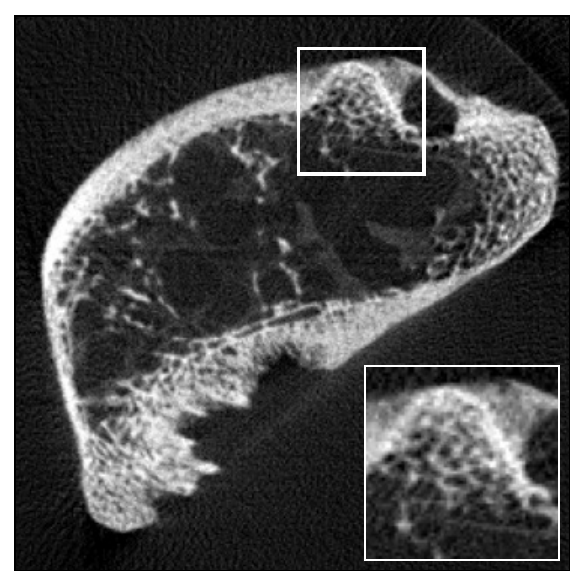}
    \end{minipage}%
    \hfill
    \begin{minipage}[c]{0.22\linewidth}
        \centering
        \includegraphics[height=\panelH]{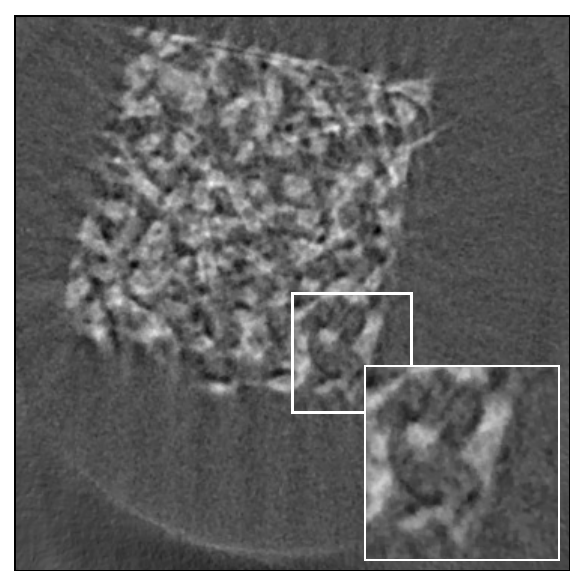}
    \end{minipage}

    \caption{Uncalibrated and calibrated reconstructions with real data.}
    \label{fig:cal_real}
\end{figure*}

With voxels, the loss landscape is piecewise constant, thus producing zero gradients 
that trap the optimizer in flat regions. The optimal solution for the geometry is also not unique. The B-spline discretization yields a 
continuously differentiable landscape, which reduces the final geometry error 
$\|\hat{\mathbf{\Theta}} - \mathbf{\Theta}^\ast\|$ by $55\%$ compared to voxels 
($0.021$ vs.\ $0.048$). Reconstructions and loss landscapes are shown in Figure~\ref{fig:cal_splines}.

\FloatBarrier
\subsection{Real Data}
We now solve the formulation~\eqref{eq:joint} for 3D real-data settings, 
each exhibiting a different type of geometric mismatch.

\paragraph{Misaligned Center of Rotation}
A micro-CT scan of a sand sample ($(768 \times 768)$ detector, $800$ projections, 
pixel size $0.127$\,mm, source-to-detector distance $765.7$\,mm, 
source-to-object distance $96.46$\,mm) as acquired in-house.\footnote{\url{https://www.epfl.ch/schools/enac/pixe/}} The 
center of rotation (COR) is parameterized as a 2D offset ($2$ unknowns), 
as illustrated in Figure~\ref{fig:calibration_scenarios}(b).

\paragraph{Source Jitter}
Each projection of a cone-beam scan of a walnut ($(972 \times 728)$ detector, $1200$ 
projections)\footnote{\url{https://www.nature.com/articles/s41597-019-0235-y}} has an independent 3D source displacement ($3 \times 1200$ 
unknowns), as illustrated in Figure~\ref{fig:calibration_scenarios}(c).

\paragraph{Angle Imprecision}
The motor angular imprecision of a cone-beam micro-CT scan of a bone sample ($(1150 \times 1150)$ detector, $361$ 
projections, source-to-object distance $210.66$\,mm, source-to-detector 
distance $553.74$\,mm)\footnote{\url{https://zenodo.org/records/6990764}} is parameterized as one unknown per projection 
($361$ unknowns), as illustrated in
Figure~\ref{fig:calibration_scenarios}(a).

\paragraph{Detector Vibrations}
A parallel-beam nano-CT scan of an NMC battery cathode 
particle was acquired at the Stanford Synchrotron Radiation Lightsource 
($(1024 \times 1024)$ detector, $180$ 
projections)\footnote{\url{https://tomobank.readthedocs.io/en/latest/source/data/docs.data.XANES.html}}. 
Each projection has an independent 2D detector offset ($(2 \times 180)$ unknowns), 
as illustrated in Figure~\ref{fig:calibration_scenarios}(c).

\begin{figure}[t]
\centering
\begin{tikzpicture}
\pgfplotsset{
    loss top/.style={
        width=0.42\columnwidth,
        height=2.2cm,
        scale only axis,
        grid=major,
        grid style={gray!15},
        ylabel style={font=\scriptsize, xshift=2pt},
        tick label style={font=\scriptsize},
        title style={font=\scriptsize, yshift=-3pt},
        xticklabels={},
        ylabel={Norm.\ loss},
        xmin=0, ymin=-0.05, ymax=1.05,
        ytick={0,0.5,1},
    },
    loss bot/.style={
        width=0.42\columnwidth,
        height=2.2cm,
        scale only axis,
        grid=major,
        grid style={gray!15},
        xlabel style={font=\scriptsize, yshift=2pt},
        ylabel style={font=\scriptsize, xshift=2pt},
        tick label style={font=\scriptsize},
        title style={font=\scriptsize, yshift=-3pt},
        xlabel={Iteration},
        ylabel={Norm.\ loss},
        xmin=0, ymin=-0.05, ymax=1.05,
        ytick={0,0.5,1},
    },
    loss right/.style={
        ylabel={},
        yticklabels={},
    },
}

\begin{axis}[loss top, name=rocks, title={Micro-CT (sand)}]
\addplot[black!35, thick, dashed]
    table[col sep=comma, header=true, x=iter, y=naive]
    {Figures/loss_norm_rocks.csv};
\addplot[black, thick]
    table[col sep=comma, header=true, x=iter, y=calib]
    {Figures/loss_norm_rocks.csv};
\end{axis}

\begin{axis}[loss top, loss right, name=walnut,
    title={Cone-beam CT (walnut)},
    at={(rocks.outer east)}, anchor=outer west, xshift=1mm]
\addplot[black!35, thick, dashed]
    table[col sep=comma, header=true, x=iter, y=naive]
    {Figures/loss_norm_walnut.csv};
\addplot[black, thick]
    table[col sep=comma, header=true, x=iter, y=calib]
    {Figures/loss_norm_walnut.csv};
\end{axis}

\begin{axis}[loss bot, name=bone, title={Micro-CT (bone)},
    at={(rocks.outer south west)}, anchor=outer north west, yshift=-1mm]
\addplot[black!35, thick, dashed]
    table[col sep=comma, header=true, x=iter, y=naive]
    {Figures/loss_norm_bone.csv};
\addplot[black, thick]
    table[col sep=comma, header=true, x=iter, y=calib]
    {Figures/loss_norm_bone.csv};
\end{axis}

\begin{axis}[loss bot, loss right, name=xrf, title={Nano-CT (NMC)},
    at={(bone.outer east)}, anchor=outer west, xshift=1mm]
\addplot[black!35, thick, dashed]
    table[col sep=comma, header=true, x=iter, y=naive]
    {Figures/loss_norm_xrf.csv};
\addplot[black, thick]
    table[col sep=comma, header=true, x=iter, y=calib]
    {Figures/loss_norm_xrf.csv};
\end{axis}

\node[anchor=north, font=\scriptsize, inner sep=0pt]
    at ($(bone.outer south)!0.5!(xrf.outer south) + (0,-3pt)$) {%
    \begin{tabular}{ll}
        \raisebox{2pt}{\tikz{\draw[black!35,thick,dashed](0,0)--(0.3,0);}}
        ~Uncalibrated &
        \raisebox{2pt}{\tikz{\draw[black,thick](0,0)--(0.3,0);}}
        ~Our method
    \end{tabular}
};

\end{tikzpicture}
\caption{Data-fidelity loss in terms of iterations for the data of Figure~\ref{fig:cal_real}.
Each panel shows the uncalibrated and calibrated curves, normalized to 
$[0, 1]$ from the joint minimum and maximum of both curves.}
\label{fig:loss_curves}
\end{figure}
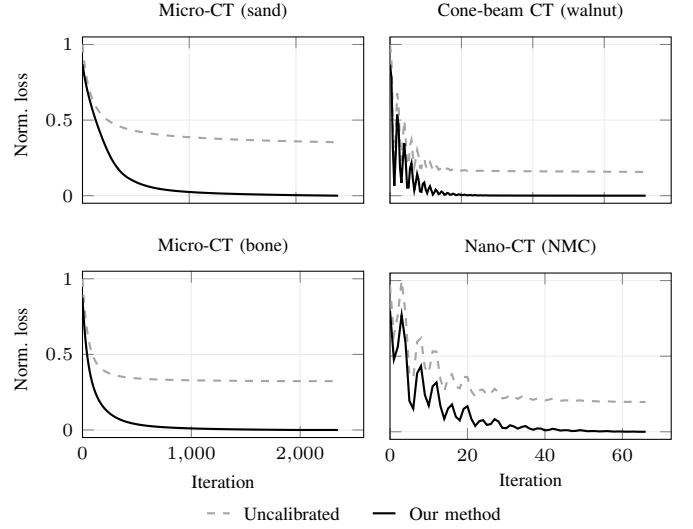

In all four cases, our calibration removes misalignment artifacts, as shown in 
Figure~\ref{fig:cal_real}. In Figure~\ref{fig:loss_curves}, we plot the loss $\mathcal{L}(\vx_k, \boldsymbol{\Theta}_k)$ at each iteration $k$ of optimization. This confirms that our method reaches a lower minimum than uncalibrated baselines.

\section{Conclusion}
\color{black}
We have presented a ray-tracing algorithm that gives access to the gradients of the 
x-ray transform with respect to arbitrary acquisition parameters. Its computational complexity is identical to that of the forward operator. The representation of the volume with B-spline basis functions yields a continuously differentiable 
forward model. Our optimization landscape is smoother than the one that results from voxel-based 
representations. We have demonstrated joint reconstruction and calibration on 
synthetic and real data across CT, micro-CT, and PET geometries. Since 
our method handles arbitrary ray geometries, it applies to any 
tomographic setup.

\section{Acknowledgments}
We acknowledge the ENAC Interdisciplinary Platform for X-
ray micro-tomography (PIXE) for providing the real micro-CT data
used in our experiments.

\begin{figure}
    \centering
    \definecolor{natred}{rgb}{0.839, 0.153, 0.157}
\definecolor{expgreen}{rgb}{0.173, 0.627, 0.173}
\begin{tikzpicture}
  \begin{loglogaxis}[
      width=0.82\linewidth, height=0.56\linewidth,
      xlabel={traversed voxels per ray},
      ylabel={GPU memory [MB]},
      xtick={64, 128, 256, 512, 1024, 2048, 4096},
      xticklabels={64, 128, 256, 512, 1024, 2048, 4096},
      x tick label style={rotate=45, anchor=east, font=\footnotesize},
      y tick label style={font=\footnotesize},
      label style={font=\small},
      legend pos=north west,
      legend cell align=left,
      legend style={draw=none, fill=none, font=\footnotesize},
      grid=both,
      grid style={gray!18},
      tick align=outside,
    ]
    \addplot[black, mark=*, thick, mark size=1.6pt, dotted]
      table[col sep=comma, x=voxels_per_ray, y=native_MB]
      {Figures/fig_autodiff_memory.csv};
    \addlegendentry{Native autodiff}
    \addplot[black, mark=*, thick, mark size=1.6pt]
      table[col sep=comma, x=voxels_per_ray, y=explicit_MB]
      {Figures/fig_autodiff_memory.csv};
    \addlegendentry{Our method}
  \end{loglogaxis}
\end{tikzpicture}

    \caption{GPU memory overhead comparison.}
    \label{fig:memory}
\end{figure}
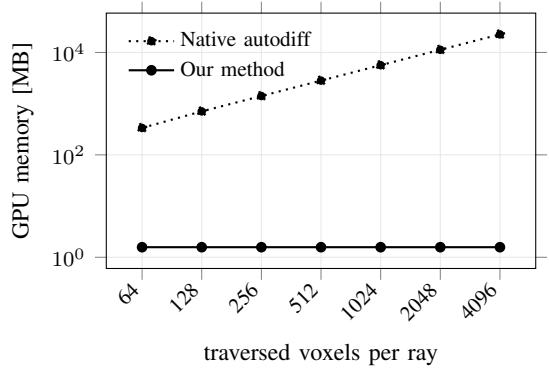

\appendix
\label{sec:appendix}
A natural alternative to the explicit passes is to differentiate through the forward traversal itself, with the automatic differentiation of Dr.Jit or PyTorch. It returns the same gradient, but reverse-mode accumulation must unroll the DDA loop and store one tape entry per traversal step. Per traversed voxel, this costs about $90\times$ more wall time, and the memory grows as $\mathcal{O}(\text{rays}\times\text{voxels})$ instead of $\mathcal{O}(\text{rays})$. It reaches $22$~GB for $4096$ rays, where the explicit pass uses under $2$MB (Figure~\ref{fig:memory}).

\FloatBarrier
\printbibliography

\end{document}